\documentclass{aa}

\usepackage{graphicx}
\usepackage{txfonts}
\usepackage{hyperref}

\begin{document}


\title{Global 3D radiation-hydrodynamical models of AGB stars\\
       with dust-driven winds}

\titlerunning{Global 3D models of AGB stars with winds} 

\author{
  Bernd Freytag
    \and
  Susanne H{\"o}fner
}

\institute{Theoretical Astrophysics,
           Department of Physics and Astronomy,
           Uppsala University,
           Box 516,
           SE-751 20 Uppsala,
           Sweden \\
           \email{Bernd.Freytag@physics.uu.se}
}

\date{\today}


\abstract
  {Convection and mass loss by stellar winds are two dynamical processes
  that shape asymptotic giant branch (AGB) stars and their evolution.
  Observations and earlier 3D models
  indicate that giant convection cells cause high-contrast surface intensity patterns, and 
  contribute to the origin of clumpy dust clouds. 
}
  {We study the formation and resulting properties of dust-driven winds from AGB stars, using new global 3D simulations. 
}
  {The dynamical stellar interiors, atmospheres, and wind acceleration zones of two M-type AGB stars were modeled with the CO5BOLD code. These first global 3D simulations are based on frequency-dependent gas opacities, and they feature time-dependent condensation and evaporation of silicate grains. 
}
  {Convection and pulsations emerge self-consistently, allowing us to derive wind
properties (e.g., mass-loss rates and outflow velocities), without relying on
parameterized descriptions of these processes.
In contrast to 1D models with purely radial pulsations, the shocks induced by
convection and pulsation in the 3D models cover large parts, but not the entirety,
of the sphere, leading to a patchy, nonspherical structure of the atmosphere.
Since dust condensation critically depends on gas density,
new dust clouds form mostly in the dense wakes of atmospheric shocks,
where the grains can grow efficiently.
The resulting clumpy distribution of newly formed dust leads to a complex 3D morphology
of the extended atmosphere and wind-acceleration zone,
with simultaneous infall and outflow regions close to the star.
Highly nonspherical isotherms and short-lived cool pockets of gas in the
stellar vicinity are prominent features. Efficient dust formation sets in closer
to the star than spherical averages of the temperature indicate, in dense
regions where grain growth rates are higher than average. This can lead to weak
outflows in situations where corresponding 1D models do not produce winds.
For stars where the overall conditions for dust formation and wind acceleration
are favorable, it is unclear whether the resulting mass-loss rates will be
higher or lower than in the 1D case. The increased efficiency of dust formation
in high-density clumps may be offset by
a low volume coverage of the forming clouds.
}
  {A global 3D approach is essential to make progress in understanding dynamical processes in AGB stars, 
  and, in particular, to solve long-standing problems regarding mass loss.
}

\keywords{
  convection --
  shock waves --
  stars: AGB and post-AGB --
  stars: atmospheres --
  stars: oscillations (including pulsations) --
  stars: mass-loss
}

\maketitle

\section{Introduction}\label{s:intro}

During the late stages of evolution on the asymptotic giant branch (AGB), low- and intermediate-mass stars are strongly affected by large-scale dynamical processes. Convection and mass loss through stellar winds influence the evolution, appearance, and final fate of these stars. They cause an enrichment of the surrounding interstellar medium with nucleosynthesis products and dust. Large-amplitude, long-period pulsations play a critical role in the formation of the dust grains that drive the massive outflows of AGB stars through radiation pressure. The pulsations trigger atmospheric shock waves, which lift gas to distances where temperatures are sufficiently low to allow for condensation of silicates and other relevant solids \citep[for a recent review on AGB mass loss, see][]{Hoefner2018A&ARv..26....1H}. 

The current theoretical picture of dynamical atmospheres and dust-driven winds of AGB stars is mainly derived from time-dependent spherically symmetric models \cite[e.g.,][]{Winters2000A&A...361..641W, Jeong2003A&A...407..191J, Hoefner2003A&A...399..589H, Hoefner2016A&A...594A.108H}. Such simu\-la\-tions describe the varying radial profiles of densities, temperatures, velocities, and dust properties, accounting for the shock waves that are triggered by pulsation and propagate outward through the atmospheres. Radiation-hydrodynamical DARWIN models, combining frequency-dependent radiative transfer with nonequilibrium dust formation, suggest that the dust-driven winds of M-type AGB stars are initiated by photon scattering on Fe-free silicate grains \citep{Hoefner2008A&A...491L...1H}. Such particles are nearly transparent at visual and near-IR wavelengths \cite[e.g.,][]{Jaeger2003A&A...408..193J, Zeidler2011A&A...526A..68Z}, resulting in significantly less radiative heating by stellar photons and, consequently, smaller condensation distances, than for their Fe-bearing counterparts \cite[e.g.,][]{Woitke2006A&A...460L...9W, Hoefner2022A&A...657A.109H}. In order to cause sufficient radiation pressure by pure scattering, however, these dust grains have to grow to rather large sizes (typical radii of about 0.1 -- 1$\,\mu$m;  comparable to wavelengths near the stellar flux maxi\-mum). The existence of such large grains in the close stellar environment is consistent with spatially resolved observations of scattered light around several nearby AGB stars \citep[e.g.,][]{Norris2012Natur.484..220N, Ohnaka2016A&A...589A..91O, Ohnaka2017A&A...597A..20O}. 

Over the past few decades, 1D dynamical models of AGB-star atmospheres and winds have given valuable insights into the critical physical processes and dust properties, as outlined above, and on the dependence of mass-loss rates on stellar parameters \cite[e.g.,][]{Wachter2002A&A...384..452W, Bladh2015A&A...575A.105B, Bladh2019A&A...626A.100B}. The computational efficiency of such simulations has made it possible to generate large grids of models, but the computational domains usually do not include the regions of the star where the pulsations are excited. The effects of stellar pulsation on the atmosphere and wind are typically introduced through variable physical conditions at the inner boundary, just below the stellar photosphere. Periodic motions simulate the radial expansion and contraction of the star, accompanied by luminosity variations, in so-called piston models. In this approach, the periods and amplitudes of the variations are free parameters that need to be constrained by observations or models of the pulsating stellar interior. 

Producing realistic theoretical models of pulsating AGB stars, however, has turned out to be a difficult task. In recent years, progress has been made regarding lower amplitude overtone pulsation by applying a linear nonadiabatic approach \citep[e.g.,][]{Wood2015MNRAS.448.3829W, Trabucchi2017ApJ...847..139T}. Lately, using 1D nonlinear pulsation models, \cite{Trabucchi2021MNRAS.500.1575T} resolved some earlier discrepancies between predicted and observed radial fundamental-mode periods of Mira variables. 

A basic problem with 1D stellar interior models is that they have to use a parameterized description of convection, which is an intrinsically 3D process. Classical descriptions such as mixing-length theory, with appropriately chosen parameters, work reasonably well in stellar evolution models. Regarding stellar pulsation, however, 1D recipes for treating convective energy transport are probably not adequate in the case of AGB stars, as they show strongly nonlinear, nonadiabatic, large-scale convective motions that couple to pulsation. Turbulent gas flows occur on scales comparable to the stellar radius, giving rise to pronounced surface intensity patterns \citep[e.g.,][]{Schwarzschild1975ApJ...195..137S, Freytag2008A&A...483..571F, Freytag2017A&A...600A.137F}. 

In recent years, with the progress in high-angular-resolution observations, it has become possible to obtain spatially resolved data for a few nearby AGB stars. Imaging of stellar surface structures at near-IR wavelengths has revealed patterns in good agreement with convective surface structures in 3D models \citep[e.g.,][]{Paladini2018Natur.553..310P}. Spatially resolved data in the submillimeter regime show complex dynamical atmospheric structures with simultaneous inward and outward motions, as well as the coexistence of hot and cold gas \citep{Khouri2016MNRAS.463L..74K, Vlemmings2017NatAs...1..848V}.  Nonspherical distributions of gas and dust in the close circumstellar environment are also seen in high-resolution images at visual and infrared wavelengths \cite[e.g.,][]{Ohnaka2016A&A...589A..91O, Stewart2016MNRAS.457.1410S, Wittkowski2017A&A...601A...3W}. Visual and near-IR scattered light images of circumstellar material have given information about the properties of dust clouds close to the star, and about the sizes of the dust grains within them \citep[e.g.,][]{Norris2012Natur.484..220N}. Temporal monitoring shows changes in both atmospheric morphology and grain sizes over the course of weeks or months \cite[e.g.,][]{Khouri2016A&A...591A..70K, Ohnaka2017A&A...597A..20O}. 

Global 3D radiation-hydrodynamical (RHD) models offer a promising way of solving existing problems with interior dynamics (convection, pulsation) and, at the same time, gaining an understanding of the complex observed atmospheric structures. The pioneering AGB "star-in-a-box" models by \cite{Freytag2008A&A...483..571F} and \cite{Freytag2017A&A...600A.137F}, building on the capability of the CO5BOLD code to cover the entire outer convective envelope and atmosphere, indeed show both large-scale convection and self-excited radial pulsations with realistic periods.
Recently, \citet{Ahmad2023A&A...669A..49A} analyzed a much larger sample of global 3D CO5BOLD models of evolved stars, finding a good agreement of the results with the observed period-luminosity relations for AGB stars.

An inhomogeneous distribution of atmospheric gas, as seen in observations, is a natural consequence of large-scale convective flows below the photosphere and the resulting network of atmospheric shock waves. The exploratory models of \cite{Freytag2008A&A...483..571F} and \cite{Hoefner2019A&A...623A.158H} indicated that the dynamical patterns in the gas will leave imprints on the dust in the close stellar environment, due to the density and temperature sensitivity of the grain growth process. However, these earlier 3D simulations did not include the effects of radiation pressure on dust, and could therefore not predict the structure of the wind formation zone. 

In this paper, we present the first global 3D RHD simulations of dust-driven winds of AGB stars, exploring the interplay of convection, pulsation, atmospheric shocks, dust formation, and wind acceleration. In addition to the newly implemented radiative pressure on dust, these new models also feature a much larger computational domain, covering the inner wind region. This allows us to follow the emerging 3D structures to a distance where the outflow is established, and to compute mass-loss rates. In Sect.~\ref{s:setup}, we give a brief overview of the basic physical assumptions and numerical methods. The results are presented in Sect.~\ref{s:results}, and compared to observations in Sect.~\ref{s:discussion}. Finally, a summary and conclusions are given in Sect.~\ref{s:conclusions}.

\section{Setup of global AGB-star models}\label{s:setup}

Below, we give a short summary of the physical and numerical properties of the CO5BOLD code, focusing on features that are relevant for the new simulations presented in this paper. More details can be found in our earlier papers on global AGB-star models \citep{Freytag2008A&A...483..571F, Freytag2017A&A...600A.137F, Hoefner2019A&A...623A.158H} and on the properties of wind-driving dust grains in M-type AGB stars \citep{Hoefner2016A&A...594A.108H}.

\subsection{General properties of the CO5BOLD models}

The CO5BOLD code \citep{Freytag2012JCP...231..919F} numerically integrates the coupled nonlinear equations of compressible hydrodynamics and nonlocal radiative energy transfer on a Cartesian grid.
The hydrodynamics scheme is based on an approximate Riemann solver of Roe-type \citep[see][]{Freytag2013MSAIS..24...26F}, modified to account for the effects of ionization and gravity. The tabulated equation of state (assuming solar abundances) takes the ionization of hydrogen and helium, and the formation of H$_2$ molecules into account. Gravitation is included as an external potential, with a general $1/r$ profile, that is smoothed in the central region of the star. In this central volume, heat is added as a constant source term, corresponding to the stellar luminosity. A drag force is active in this core region only, to prevent dipolar flows traversing the entire star. All outer boundaries are open for the flow of matter and for radiation \citep[see][for some details about boundary conditions in CO5BOLD]{Freytag2017MmSAI..88...12F}.

\subsection{Computational domain and radiative transfer}\label{s:box_radtra}

\begin{table*}[htb]
  \begin{center}
  \caption{Basic model parameters and derived quantities. 
  \label{t:ModelParam}}
  \begin{tabular}{lrrrrrrrrrrrr}
\hline\hline
model & $M_\star$ & $M_\mathrm{env}$ & $L_\star$ & $n_x^3$ & $x_\mathrm{outerbox}$ & $x_\mathrm{innerbox}$ & $C_{T\mathrm{fac}}$ & $t_\mathrm{avg}$ & $R_{\star,s_\mathrm{min}}$ & $T_{\mathrm{eff},s_\mathrm{min}}$ & $\log g_{s_\mathrm{min}}$ & $P_\mathrm{puls}$ \\
  & ($M_\sun$) & ($M_\star$) & ($L_\sun$) &   & ($R_\sun$) & ($R_\sun$) &  & (yr) & ($R_\sun$) & (K) & cgs & (d) \\
\hline
st28gm06n050 &  1.0 & 0.182 &  7049 & 599$^3$ & 4858 & 2340 & 0.75 & 54.61 &  351 & 2823 & -0.656 & 510 \\
st28gm06n052 &  1.0 & 0.181 &  7030 & 679$^3$ & 6386 & 2640 & 0.77 & 57.78 &  355 & 2806 & -0.665 & 545 \\
st28gm05n033 &  1.5 & 0.298 &  6702 & 559$^3$ & 3454 & 1581 & 0.72 & 27.70 &  304 & 2993 & -0.358 & 297 \\
\hline
  \end{tabular}
 \end{center}
\tablefoot{The table shows the model name;
the mass $M_\star$, used for the external potential;
the envelope mass $M_\mathrm{env}$, derived from integrating the mass density of all grid cells within the computational box; 
the average emitted luminosity $L_\star$; 
the model grid dimensions $n_x^3$;
the edge length of the entire cubical computational box $x_\mathrm{outerbox}$;
the edge length of the inner box with detailed radiation transport $x_\mathrm{innerbox}$;
the adjustable temperature-reduction factor $C_{T\mathrm{fac}}$ in the outer layers;
the time $t_\mathrm{avg}$, used for averaging the remaining quantities in this table; 
the radius $R_{\star,s_\mathrm{min}}$ at the point with minimum entropy;
the effective temperature $T_{\mathrm{eff},s_\mathrm{min}}$ at minimum entropy;
the logarithm of the surface gravity $\log g_{s_\mathrm{min}}$ at minimum entropy;
and the pulsation period $P_\mathrm{puls}$.
}
\end{table*}

A global model describing the flow of gas and radiation from the convective stellar interior, through the dynamical atmosphere, to the dust formation region and wind-acceleration zone, needs to cover a wide range of physical conditions on different spatial and temporal scales. To keep computation times at a reasonable level, the spatial domain of the new global models is divided into two regions. An inner box, covering the star and its immediate surroundings, features a more detailed description of radiative transfer, which is critical for the modeling of temperature structures (see below) and a higher spatial resolution. It is surrounded by a larger outer box with a simplified radiative transfer and a coarser grid, beyond the main grain growth region, where the wind has been initiated and dynamical structures tend to be larger. Regarding hydrodynamics, there is no difference between the two boxes, except for the grid spacing. The inner box consists of cubical cells with a constant size. In the outer box, going outward, cells become incrementally larger along the axis directions (typically by a few percent per step) and only those along the space diagonals retain a cubical shape, while their size increases.  

In the inner box, the nonlocal radiative energy transfer is solved with a short-characteristics scheme. Most of our earlier global 3D RHD simulations of AGB stars used tabulated gray gas opacities, which is sufficient for studies of interior properties such as convection and pulsations, and to give a qualitative picture of the shock-dominated atmospheric dynamics \citep[see][]{Freytag2017A&A...600A.137F}. Simulations including dust formation, however, require a refined modeling of the atmospheric temperature structure, 
which sets thresholds for the onset of dust condensation and evaporation. This can be achieved by using frequency-dependent data, as discussed in \cite{Hoefner2019A&A...623A.158H}. In the new models of dust-driven winds presented in this paper, we use a similar approach with three frequency bins, but we have redone the iterative binning procedure for a larger region of the representative pressure-temperature structure, due to the larger computational box used here. The tables for atmospheric gas opacities used in the present models are based on COMA data \citep[see][]{Aringer2000DissAri, Aringer2016MNRAS.457.3611A}, extended with OPAL data at temperatures above approximately 12\,000\,K. Scattering is treated as true absorption, that is to say, the scattering opacity is added to the absorption opacity, so that the source function can be computed from the local temperature alone.

In the outer box, where the outflow is established and further grain growth is of minor importance, temperature plays a less critical role. Therefore, we use an approximate description, sufficient for the purpose of hydrodynamics, which is computationally much less costly than solving radiative transfer.  
The average radial temperature profile is assumed to be set by the radiative flux from the stellar surface, geometrically diluted with distance, which can be written as $T(r) \propto (L/r^2)^{1/4}$, where $L$ is the mean stellar luminosity and $r$ the radial distance from the center.
To minimize the jump in temperature from the inner box
(with detailed radiative transfer, as described above), an adjustable factor $C_{T\mathrm{fac}}$ (see Table\,\ref{t:ModelParam}) is included in the description of $T(r)$. In each time step, the internal energy at each grid point in the outer layers is adjusted, so that the actual temperature relaxes toward the approximate target temperature on a small but finite timescale (typically $10^4$\,s).

\subsection{Dust formation}\label{s:dust_spec}

In the new global models with dust-driven winds, we use a time-dependent kinetic treatment of silicate formation and destruction, adapted from the DARWIN models, as described in detail by \cite{Hoefner2016A&A...594A.108H}. The dust grains grow by the addition of abundant atoms and molecules from the gas phase, and they may shrink due to thermal evaporation from the grain surface. The growth of grains is triggered by the temperature falling below a critical value \citep[depending on gas density; see Fig.\,1 in][]{Hoefner2016A&A...594A.108H}. Conversely, when the temperature rises above this value, the grains start to shrink due to evaporation from the surface. At the relatively low densities in the stellar atmosphere, grain growth typically occurs on timescales that are comparable to those of gas dynamics and radiative flux variations. Grain growth and evaporation may, therefore, proceed far from equilibrium, making a time-dependent treatment necessary. 

The condensation of wind-driving olivine-type silicate grains is assumed to proceed according to the net reaction
\begin{equation}\label{e_path_ol}
  {\rm 2 \, Mg + SiO + 3 \, H_2 O }  \,\,  \longrightarrow  \,\,  {\rm Mg_2 SiO_4 + 3 \, H_2 }
    \enspace .
\end{equation}
In principle, olivine-type silicates can be considered as a solid solution of Mg$_2$SiO$_4$ and Fe$_2$SiO$_4$, with a variable Fe/Mg ratio. However, as discussed by \cite{Hoefner2022A&A...657A.109H}, the inclusion of Fe atoms in the growing grains is a secondary process, taking place after the wind has been triggered by Fe-free silicate dust. The Fe/Mg ratio remains low, due to a self-regulating feedback via the grain temperature, and due to rapidly falling densities in the outflow. The resulting effects on wind dynamics are small, and therefore we only consider the condensation of Fe-free silicates (Mg$_2$SiO$_4$) in the exploratory 3D wind models presented here.  

It should be noted that the kinetic treatment of grain growth used here does not describe nucleation, that is, the formation of the very first solid seed particles. Since nucleation rates and the chemical composition of seed particles in M-type AGB stars are still a matter of debate, the abundance of seed particles relative to hydrogen is treated as an input parameter \citep[for a more detailed discussion see][]{Hoefner2016A&A...594A.108H}. It is assumed that these seed particles are readily available whenever conditions permit the condensation of silicate dust. They are tiny compared to the resulting dust grains, and they have no effect other than providing an initial condensation surface for grain growth.

\subsection{Radiation pressure on dust}\label{s:dust_rp}

In contrast to the pre-tabulated gas opacities (see Sect.\,\ref{s:box_radtra}), the dust opacities that cause radiative acceleration are calculated during the simulations, using the current grain radii that result from the equations describing dust condensation and evaporation (see Sect.\,\ref{s:dust_spec}). The total opacity of grains with radius $a_{\rm gr}$ in a volume element (cross section per volume) can be expressed as 
\begin{equation}
  \kappa_{\rm acc} (a_{\rm gr},\lambda) = \pi a_{\rm gr}^2 \, Q_{\rm acc} (a_{\rm gr},\lambda) \, n_{\rm d}
    \enspace ,
\end{equation}
where $n_{\rm d}$ is the corresponding number density of grains and $\lambda$ represents the wavelength. The efficiency factor $Q_{\rm acc} (a_{\rm gr},\lambda)$, defined as the ratio of radiative to geometrical cross section of a grain, contains contributions from true absorption and scattering:%
\begin{equation}
  Q_{\rm acc}  =  Q_{\rm abs} + ( 1 -  g_{\rm sca} ) \, Q_{\rm sca}
    \enspace ,
\end{equation}
where $g_{\rm sca}$ is an asymmetry factor describing deviations from isotropic scattering. 

In principle, the efficiency factors and $g_{\rm sca}$ can be computed using Mie theory. In the exploratory 3D wind models presented here, however, we use a simplified description of $Q_{\rm acc}$, retaining the critical dependence on grain size, but treating the dependence on wavelength in an approximate way. We assume that the flux mean of the opacity determining radiative acceleration 
can be replaced by a monochromatic value at a wavelength close to the flux maximum, $\kappa_{\rm acc} (a_{\rm gr}, \lambda_{\rm max})$, where $\lambda_{\rm max} \approx 1 \mu$m, and we use a simple analytical approximation for $Q_{\rm acc}$ as a function of grain radius:%
\begin{equation}
   Q_{\rm acc} (a_{\rm gr}, \lambda_{\rm max}) =
   \min \left[ \left( q_1 \left(\frac{a_{\rm gr}}{\lambda_{\rm max}}\right)
                    + q_4 \left(\frac{a_{\rm gr}}{\lambda_{\rm max}}\right)^4
               \right) ,
               q_{\rm max}  \right]
    \enspace ,
\end{equation}
introduced by \cite{Hoefner2008PhST..133a4007H}. The term linear in grain radius represents true absorption for grains smaller than the wavelength under consideration, while scattering in that regime is accounted for by the term depending on the fourth power of the grain radius, which dominates for near-transparent Fe-free silicate grains. The fact that the efficiency factor approaches a constant value for grains much larger than the wavelength is taken into consideration by setting a maximum value $q_{\rm max}$. Choosing $q_1 = 2.0 \cdot 10^{-3}$, $q_4 = 4.68 \cdot 10^{2}$ and $q_{\rm max} = 1.0$ at $\lambda_{\rm max} = 1.08\,\mu$m gives a description that stays reasonably close to $Q_{\rm acc}$ computed from full Mie theory, considering the level of the other approximations in the radiative transfer, and the fact that $Q_{\rm acc}$ varies over several orders of magnitude within the grain-size range of interest \citep[see Figs.\,3 and 4 in][]{Hoefner2008PhST..133a4007H}. 

To compute the radiative force acting on the dust grains, the opacity $\kappa_{\rm acc}$ has to be multiplied with a factor describing the photon flux from the star. In an optically thin environment, this corresponds to a geometrically diluted flux from the stellar surface, decreasing with radial distance roughly as $1/r^2$ (assuming that the star can be approximated by a point source). However, if the optical depth of the circumstellar dust is not negligible, leading to a weakening of the local photon flux, this effect has to be taken into account, which is done by introducing a factor $( 1 \, - \, \exp\,(-d\tau) ),$ where $d\tau$ is the local optical depth.
We chose $d\tau=0.02 r \kappa_{\rm acc}$.
This situation can arise during dust formation in small high-density regions in the wake of shocks, corresponding to typical length scales of a few percent of the radial distance $r$. In practice, however, the optical depth effects are usually negligible. 

Currently, radiation pressure is the only dust opacity effect taken into account. The Fe-free silicate grains are very transparent at near-IR wavelengths, where the stellar flux has its maximum, and the corresponding absorption coefficients are neglected when solving radiative energy transfer in the inner box to obtain the temperature structure. The dust grains are assumed to have the same temperature as the gas, which is mainly relevant for the thresholds for condensation and evaporation. This approximation is in line with the other simplifications concerning the dust component.

\subsection{Input parameters and resulting model properties}\label{s:modpar}

\begin{figure}[hbtp]
\vspace{-0.4mm}
\includegraphics[width=8.8cm]{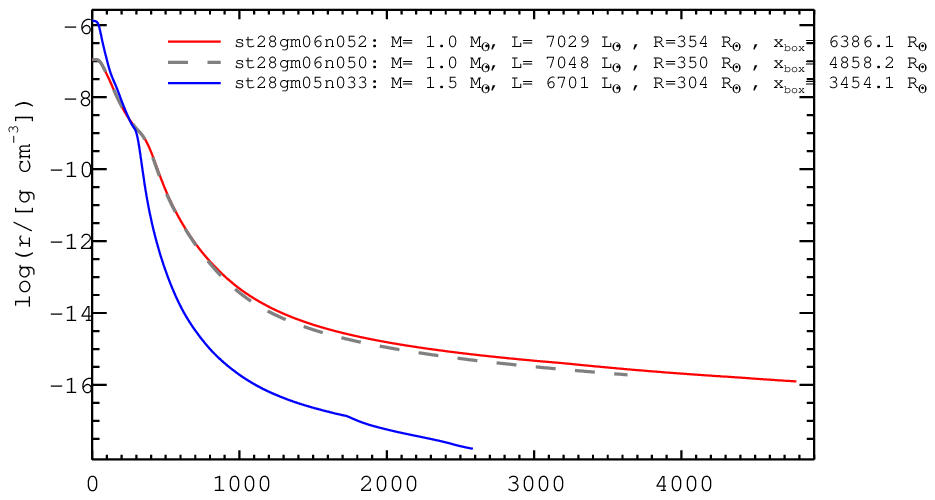}\vspace{-1.8mm}
\includegraphics[width=8.8cm]{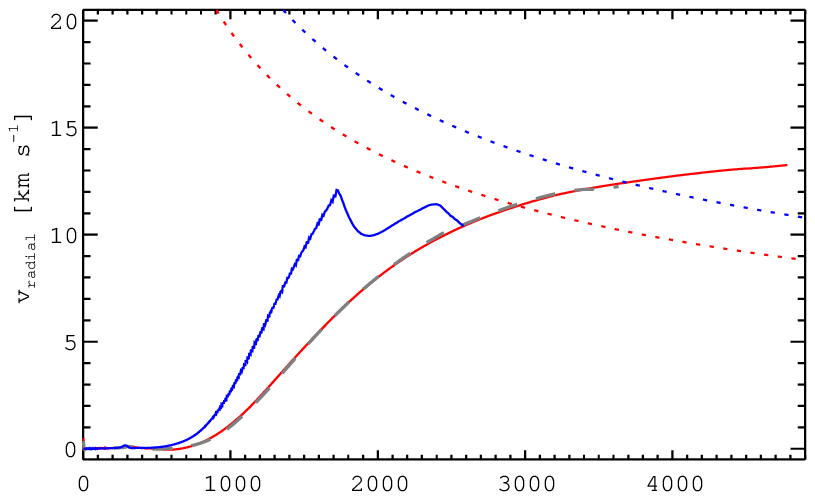}\vspace{-1.8mm}
\includegraphics[width=8.8cm]{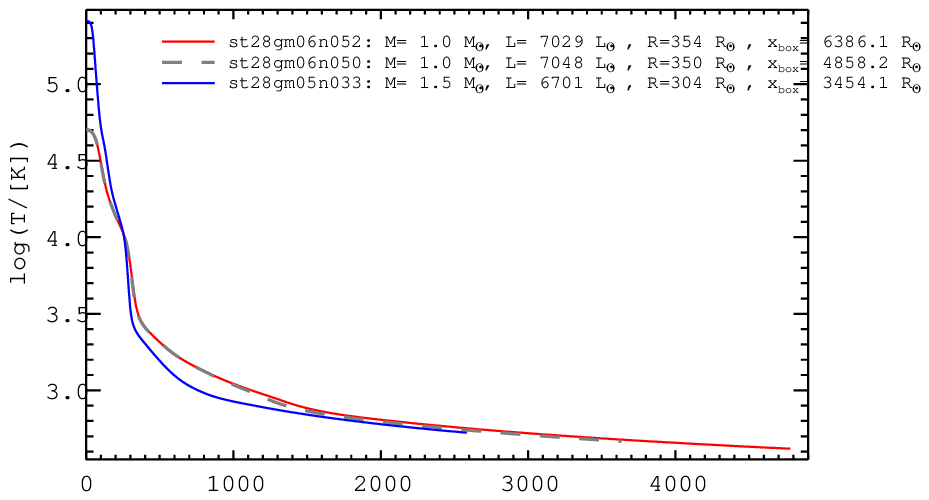}\vspace{-1.8mm}
\includegraphics[width=8.8cm]{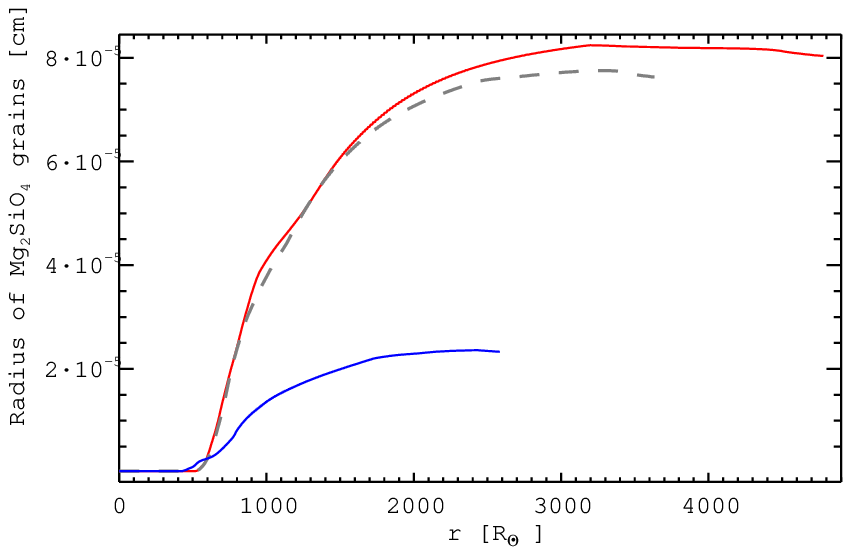}
\caption{Mean radial structures of the
models
  st28gm06n050 
    (dashed black curves), 
  st28gm06n052 
    (red curves), and 
  st28gm05n033 
    (blue curves). 
Shown are gas density, radial velocity, temperature, and silicate grain radius, averaged over spherical shells and time, and plotted against the distance from the stellar center. 
Averages are not only taken over spheres that fit completely into the cubical computational box, but also over partial spheres somewhat beyond, up to $1.5\times x_\mathrm{outerbox}/2$, omitting the regions close to the corners of the cube.
The dotted red and blue curves in the radial-velocity panel represent the escape velocity as a function of distance for stellar masses of 1\,$M_\sun$ and 1.5\,$M_\sun$, respectively. For details, see Sects.\,\ref{s:modpar} and \ref{s:results}.
\label{f:st28gm06n045_TimeAvgx}}
\end{figure}

In this paper we present a selection of new 3D AGB-star models with dust-driven winds, demonstrating the effects of stellar and numerical parameters (see Table\,\ref{t:ModelParam}). Models st28gm06n050 and st28gm06n052 are similar, except for the sizes of the inner and outer computational boxes, and the adjustable factor of the target temperature $C_{T\mathrm{fac}}$ in the outer box, which ensures a smooth transition in temperature from the domain of detailed radiative transfer to the outer box (see Sect.\,\ref{s:box_radtra}). The mean radial structures of models st28gm06n050 and st28gm06n052, averaged over spherical shells and time, are shown in Fig.\,\ref{f:st28gm06n045_TimeAvgx} (dashed black and solid red curves), demonstrating that box size and the simplified treatment of the radiative transfer in the outer box only have a minor effect on the resulting wind dynamics. In particular, the mean radial-velocity profiles are almost identical for the two models. In the following, we focus our discussion on model st28gm06n052, which has both a larger inner and outer box, and therefore allows us to follow the development of 3D structures in more detail, and to larger distances from the star. The blue curves in Fig.\,\ref{f:st28gm06n045_TimeAvgx} show the mean radial structure of model st28gm05n033, which has a higher mass, a smaller radius, and a higher effective temperature, compared to models st28gm06n050 and st28gm06n052. This leads to less favorable conditions for dust formation and wind acceleration, as is discussed in Sect.\,\ref{s:results}. 

Table\,\ref{t:ModelParam} summarizes the basic parameters and resulting global properties of the models. While the stellar mass $M_\star$ (controlling the gravitational potential), as well as the resolution and the extent of the numerical grid, are pre-chosen fixed parameters, other model properties are determined after a simulation is finished. The envelope mass $M_\mathrm{env}$ is calculated from the integrated density of all grid cells, averaged over time.
We assume that the difference in mass is located in the compact, unresolved stellar core.
The listed stellar luminosity is a time average of the total luminosity emitted at the surface (very similar to the inserted luminosity of 7000\,$L_\sun$ in the core). 

The stellar radius is more difficult to determine and is less well defined due to the complex morphology of the extended atmosphere. Here, we use a value corresponding to the point of minimum entropy near the photosphere, which has turned out to be a good choice in connection with the analysis of pulsation properties \citep{Ahmad2023A&A...669A..49A}.
We note that this definition is different from the one used in earlier papers \citep[e.g.,][]{Freytag2017A&A...600A.137F, Hoefner2019A&A...623A.158H}, where the radius was chosen as the point $R_\star$ where the spherically and temporally averaged temperature and luminosity fulfill $\langle L \rangle_{\Omega,t}$\,=$\,4\pi\sigma R_\star^2 \langle T \rangle_{\Omega,t}^4$. For the less massive and cooler models st28gm06n050 and st28gm06n052, the latter definition leads to values that are about 10\% larger, while both definitions give very similar values for model st28gm05n033, which has a more compact atmosphere (see Fig.\,\ref{f:st28gm06n045_TimeAvgx}, showing the mean radial density and temperature structures). 

In contrast to our earlier 3D simulations of AGB stars, the new models presented here predict mass-loss rates and properties of wind-driving dust grains (see Table\,\ref{t:mod}). The dust properties are a direct result of solving the equations describing grain growth and evaporation (see Sect.\,\ref{s:dust_spec}). The mass-loss rate of a model is computed by averaging the mass flux across spherical shells and time.

%
\begin{table}
\caption{\label{t:mod}Dust and wind properties.}
\centering
\begin{tabular}{lcccc}
\hline\hline
  model & $n_d/n_{\rm H}$ & $\dot{M}$    & $f_{\rm Mg}$ & $a_{gr}$ \\
        &               & ($M_\odot$/yr) &           & ($\mu$m) \\ 
\hline
  st28gm06n052 & $3 \cdot 10^{-16}$ & $5 \cdot 10^{-6}$ & 0.5  & 0.8 \\
  st28gm05n033 & $3 \cdot 10^{-15}$ & $5 \cdot 10^{-8}$ & 0.15 & 0.2 \\
\hline
\end{tabular}
\tablefoot{Listed here are the assumed seed particle abundance $n_d/n_{\rm H}$, and the resulting temporal means of the mass-loss rate $\dot{M}$, the fraction of Mg condensed into grains $f_{\rm Mg}$, and the grain radius $a_{gr}$ at the outer boundary. 
When forming Mg$_2$SiO$_4$ grains in a gas of solar composition, the abundance of Mg is the limiting factor, since that element will be used up first. In the models described here, however, $f_{\rm Mg}$ is well below its maximum value of 1. 
}
\end{table}

\begin{figure*}[hbtp]
\begin{center}
\hspace*{0.9cm}\includegraphics[width=15.3cm]{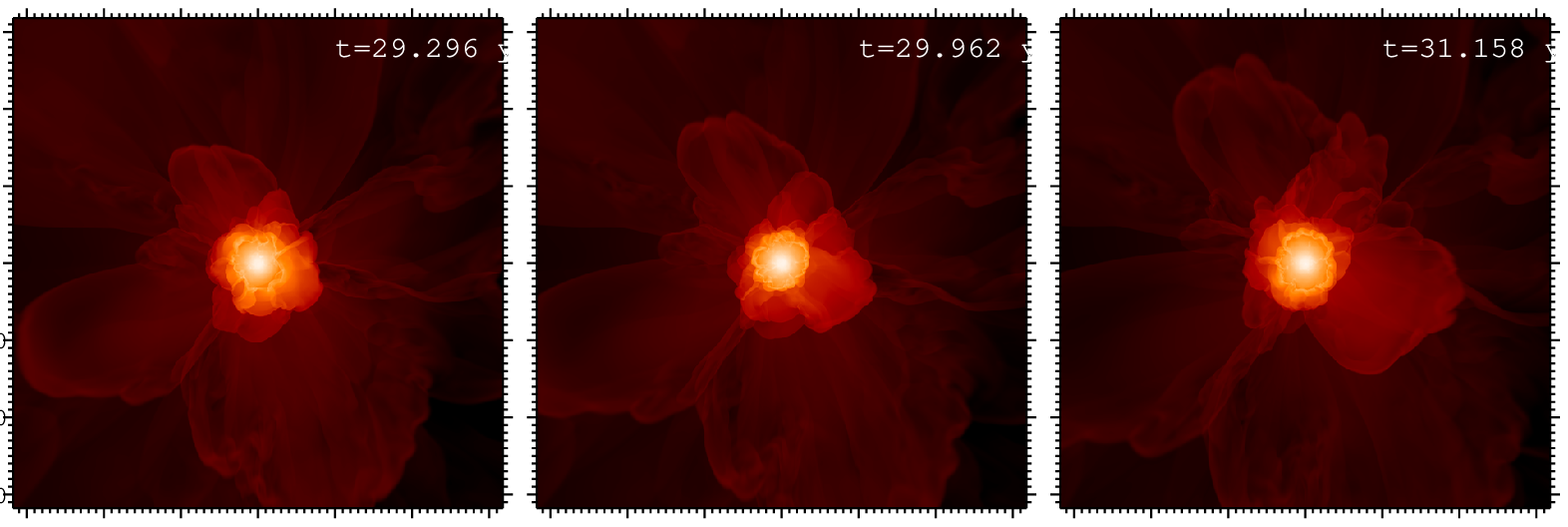}\includegraphics[width=1.9cm]{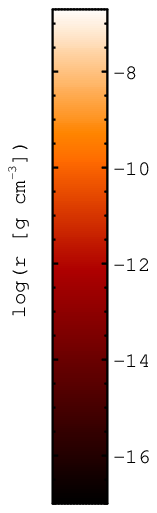}
\hspace*{0.9cm}\includegraphics[width=15.3cm]{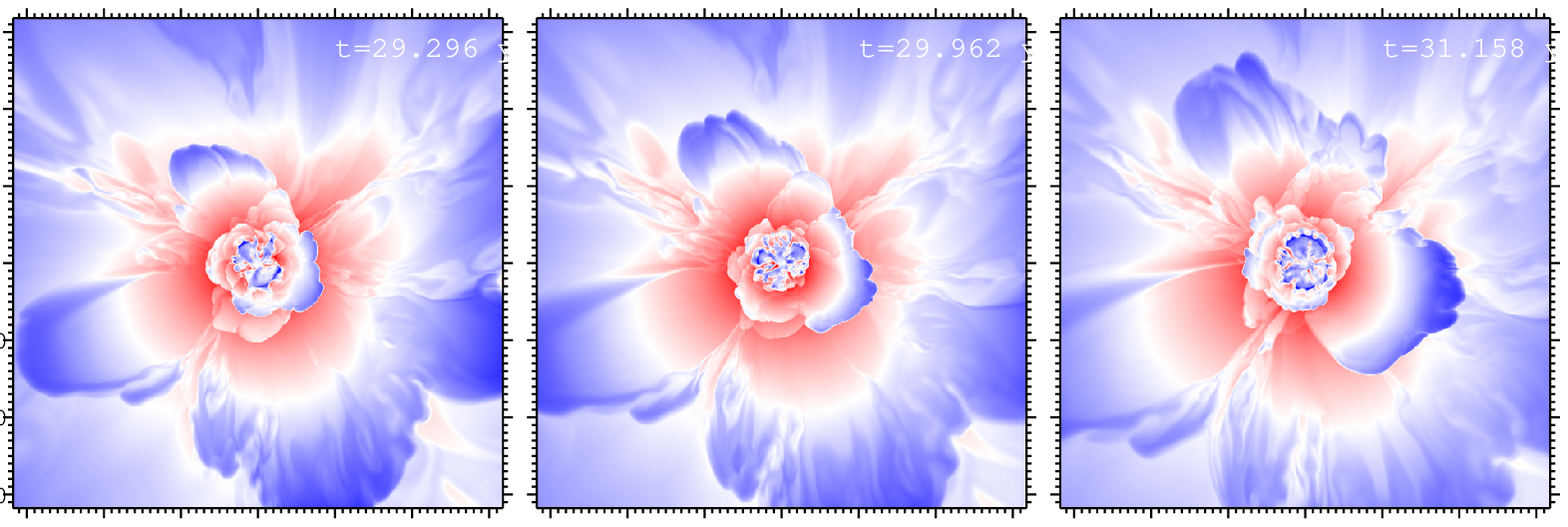}\includegraphics[width=1.9cm]{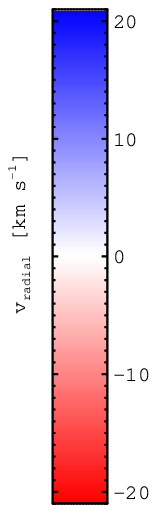}
\hspace*{0.9cm}\includegraphics[width=15.3cm]{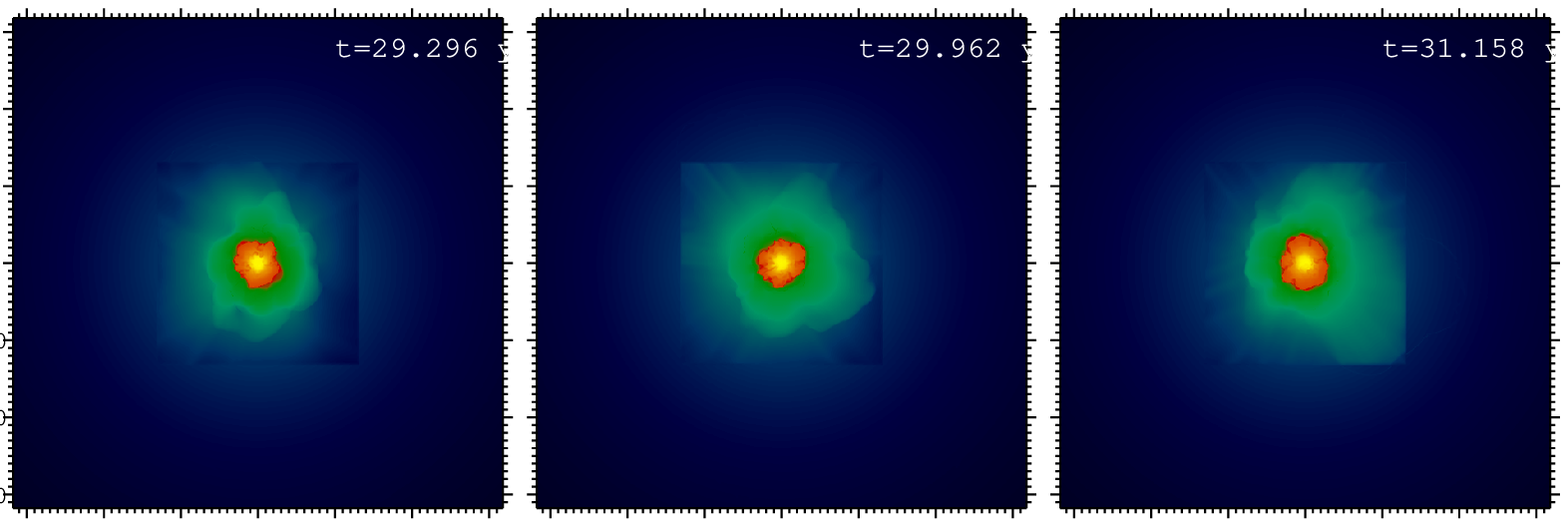}\includegraphics[width=1.9cm]{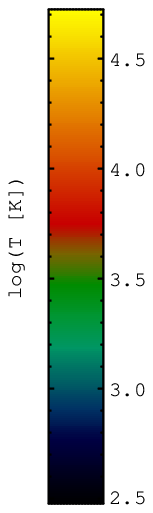}
\hspace*{0.9cm}\includegraphics[width=15.3cm]{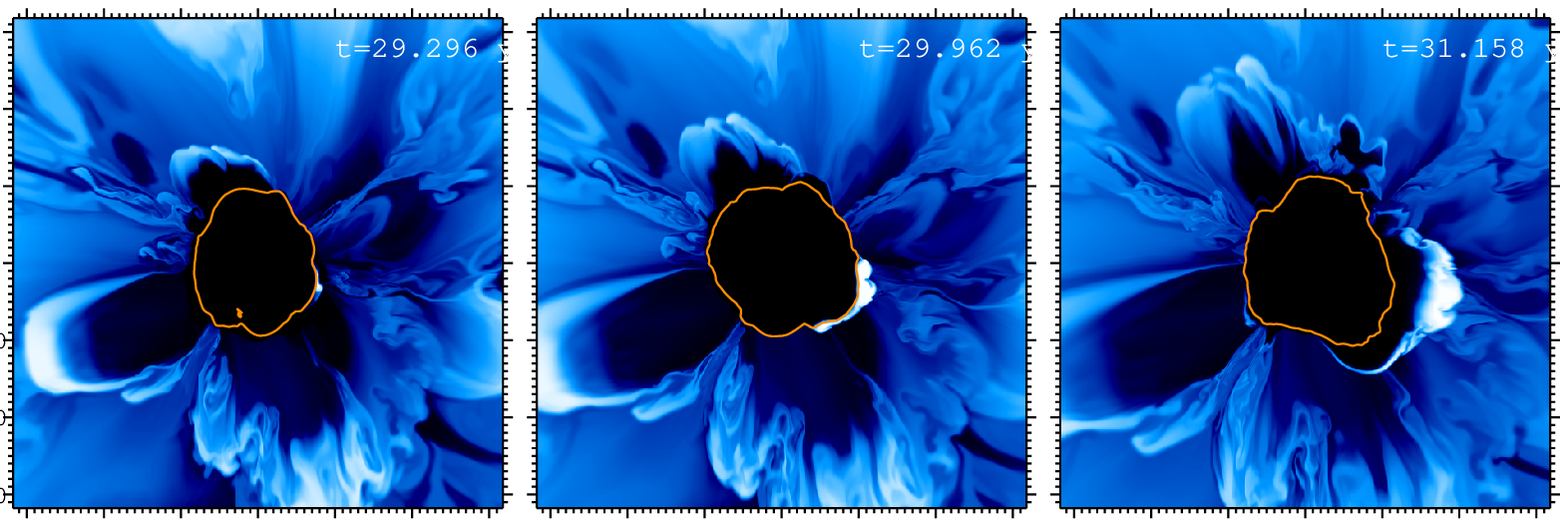}\includegraphics[width=1.9cm]{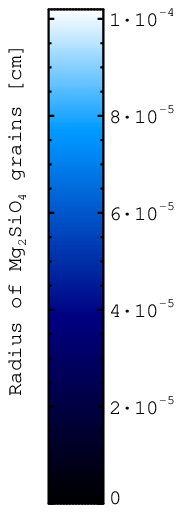}
\end{center}
\caption{Time sequences of density, radial velocity, temperature, and silicate grain radius 
  for a slice through the center of the large 1\,$M_\sun$~model st28gm06n052.
  The snapshots are about 8 and 14 months apart, respectively (see the counter at the top of the panels).
  The orange line in the bottom panels indicates an isotherm of 1150\,K.
\vspace{-1mm}
\label{f:st28gm06n052_0233896_QuSeq1}}
\end{figure*}

\begin{figure*}[hbtp]
\begin{center}
\hspace*{0.9cm}\includegraphics[width=15.3cm]{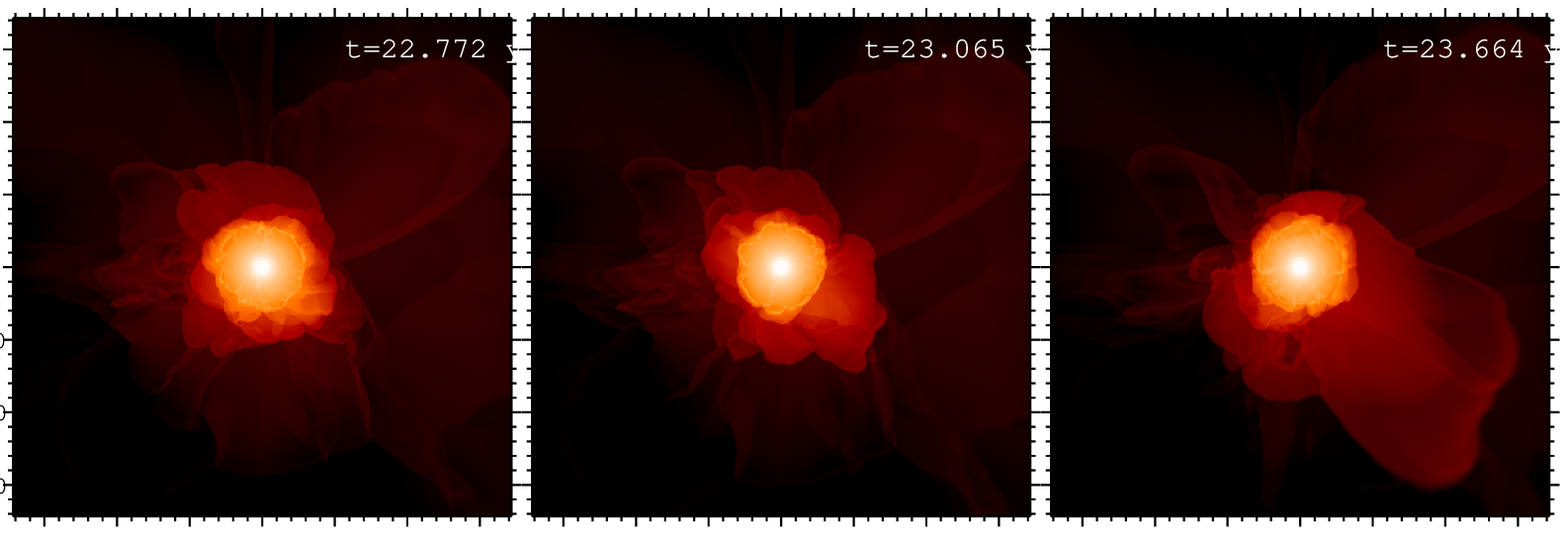}\includegraphics[width=1.9125cm]{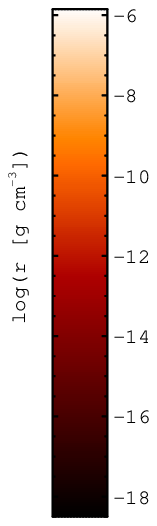}
\hspace*{0.9cm}\includegraphics[width=15.3cm]{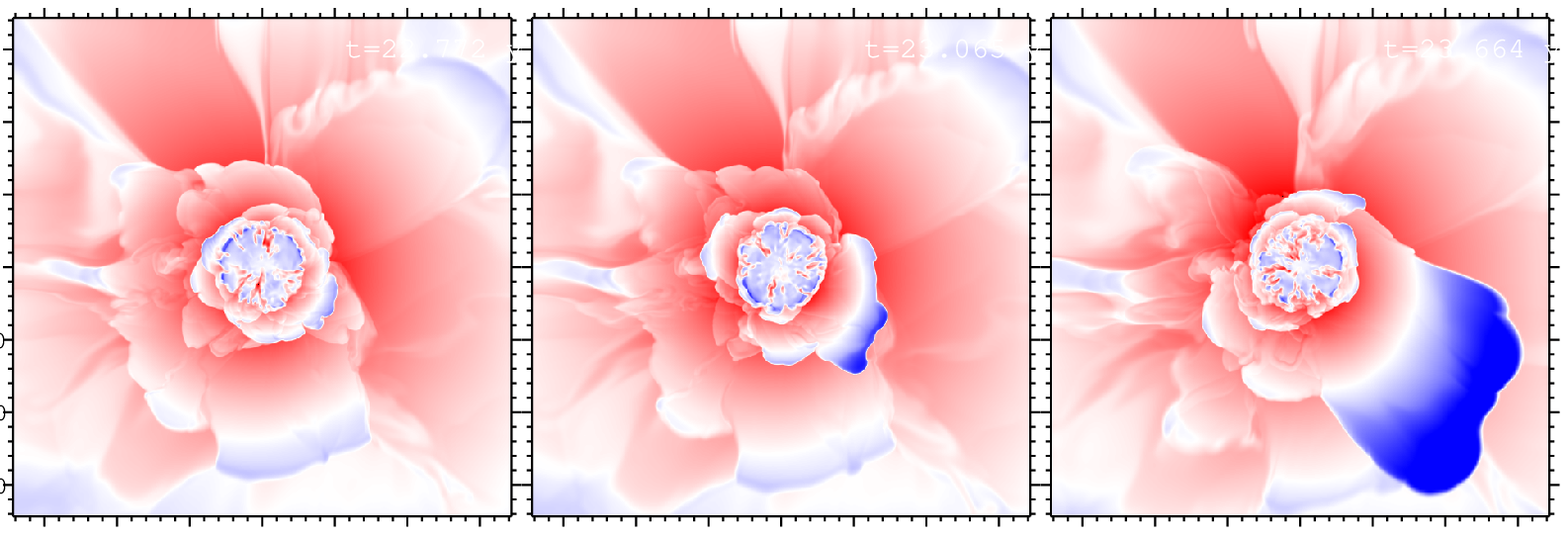}\includegraphics[width=1.9125cm]{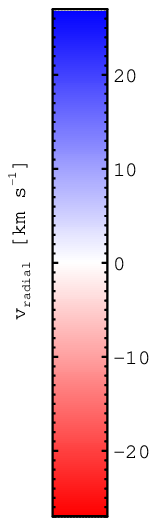}
\hspace*{0.9cm}\includegraphics[width=15.3cm]{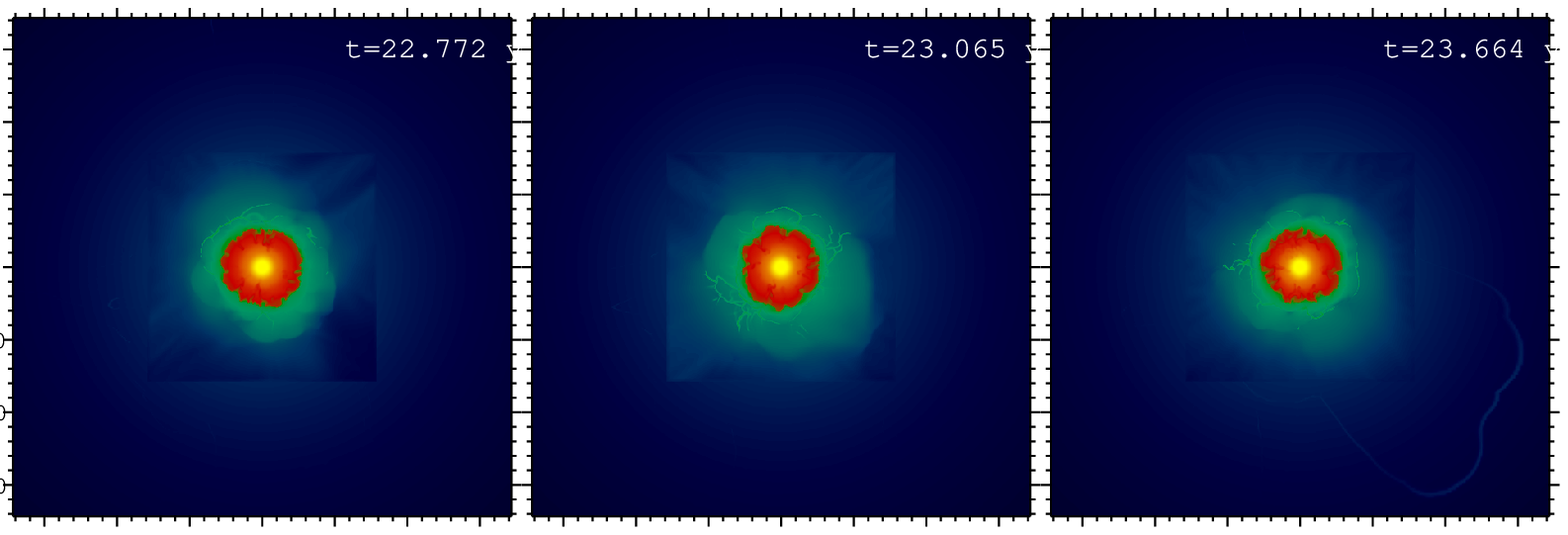}\includegraphics[width=1.9125cm]{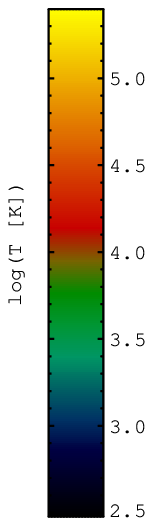}
\hspace*{0.9cm}\includegraphics[width=15.3cm]{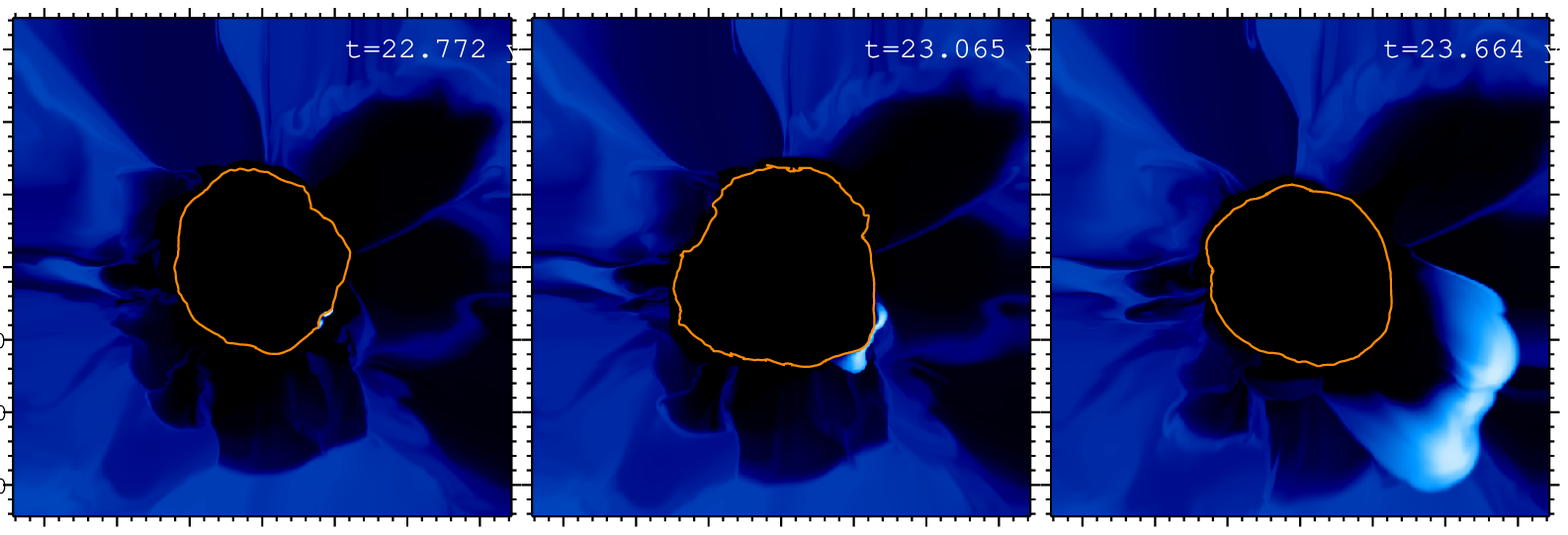}\includegraphics[width=1.9125cm]{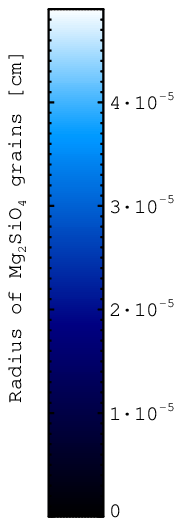} 
\end{center}
\caption{Time sequences of density, radial velocity, temperature, and silicate grain radius 
  for a slice through the center of the 1.5\,$M_\sun$~model st28gm05n033. 
  The snapshots are about 3.5 and 7 months apart, respectively (see the counter at the top of the panels).
  The orange line in the bottom panels indicates an isotherm of 1150\,K.
\vspace{-1mm}
\label{f:st28gm05n033_0375208_QuSeq1}}
\end{figure*}

\section{Results}\label{s:results}

In this section, we describe in detail the emergence and evolution of atmospheric and circumstellar structures triggered by large-scale convective flows and stellar pulsation, and study the persistence of inhomogeneities in the gas and dust distributions during the onset of dust-driven outflows. In the following, we focus on the description of these phenomena in model st28gm06n052, using the more massive and hotter model st28gm05n33 mainly to illustrate the influence of stellar parameters.

\subsection{Formation and evolution of dust clouds}

Figure \ref{f:st28gm06n052_0233896_QuSeq1} shows the evolving structures in model st28gm06n052 for slices through the center of the star, at three instants of time (left to right). The top row of panels shows the gas density, illustrating the strong contrast between the high values in the stellar interior (bright colors at the center of the image) and the thin circumstellar material, with a steep transition region in the inner atmosphere (it is important to note the logarithmic scale).

In the second row of panels, showing radial velocity, these different zones can be distinguished by their dynamical behavior. The stellar interior (inside a radius of about 350\,$R_{\odot}$ from the center) is dominated by large-scale convective flows, corresponding to variable blue and red areas that represent outflows and inflows, respectively. In the atmosphere, the convective flows, together with radial pulsations, trigger strong, outward-propagating shock waves, seen as dark blue arc-like structures
in the 2D slices. Beside these shock-accelerated regions, there are large areas where gas is falling back toward the star (apparent in red), since the inner atmosphere is gravitationally bound. The wind-acceleration region corresponds to the blue region in the outer parts of the model, indicating the outflow of matter. As discussed above, the stellar wind is driven by radiation pressure on dust. The grains collide with the surrounding gas particles and transfer outward-directed momentum, thereby initiating the wind.

The bottom row of panels in Fig.~\ref{f:st28gm06n052_0233896_QuSeq1} shows the silicate grain radius, with brighter colors indicating larger grains. The spatial patterns in grain size reflect the effects of density and temperature on the condensation process. Higher densities lead to faster, more efficient grain growth, and the brightest areas (largest grains) are found in the wakes of shock waves, which compress the gas. The central dark zone indicates the region where dust formation is prevented by high temperatures (see panels in row 3). The over-plotted line in the grain-size panels represents an isotherm at 1150\,K, corresponding roughly to the condensation temperature of the silicate grains.

In the bottom right quadrant of the images, a new dust cloud (bright area in grain-size plots) is forming in the wake of an outward-propagating shock wave, which compresses the gas (blue arc in the radial velocity plots and brighter orange in the density plots). Initially appearing as a small bright region in the grain radius plot (bottom row, left panel), the cloud quickly grows in size, with its inner edge defined by the condensation temperature (middle panel), before being driven outward by radiation pressure (right panel). In the bottom left quadrant, an existing cloud that was formed earlier in a similar way, is accelerated outward by radiation pressure, leaving the model domain. Another outward-moving cloud can bee seen in the top left quadrant. The fact that the series of snapshots (covering about 22 months in total) shows several distinct dust clouds indicates that efficient dust formation events, triggered by shock waves, occur frequently. It should, however, be mentioned here that not all dust clouds survive and lead to an outflow of matter. In some cases, the variations in the radiative flux and the resulting changes in atmospheric temperatures lead to dust evaporation, before the cloud is accelerated outward. This is consistent with the coexistence of inward and outward moving gas in the atmosphere, mentioned above.

\subsection{Overall morphology and wind properties}

The dependence of grain growth rates on density gives the stellar wind a patchy, time-variable start. As clumps of dusty gas move outward and merge into a general outflow, some degree of inhomogeneity is preserved. Looking at Fig.~\ref{f:st28gm06n052_0233896_QuSeq1}, there are clear signs of a stellar wind in the outer regions of model st28gm06n052, with a dominant direction of the flow velocity away from the star. However, the density and dust-grain size still show imprints of the atmospheric shock waves that triggered dust condensation, and thereby the onset of the wind. 

The radial structure of model st28gm06n052, averaged over spherical shells and time, is shown in Fig.\,\ref{f:st28gm06n045_TimeAvgx} (red curves). In the plots of the mean radial velocity, the time-dependent dynamics of the stellar interior and atmosphere averages out, and the dominant flow direction of the wind is clearly visible. After an initial phase of fast grain growth, dust condensation slows down, as density drops rapidly in the accelerating wind. At about $3000\,R_\sun$ (corresponding to about eight stellar radii), the average radial velocity of the flow exceeds the local escape velocity (dotted red line). When the outflow approaches a constant velocity further out, the density profile changes from a steep decline to the familiar $1/r^2$ profile of a spherical wind with constant velocity. 

In Fig.\,\ref{f:st28gm06n045_TimeAvgx}, the time-averaged radial structures of models with different stellar parameters are compared. It is evident that model st28gm05n033 (blue curves), representing a more massive, hotter star with a smaller radius, shows a much steeper atmospheric density decrease. At a distance where temperatures are low enough for dust condensation, the difference in density compared to model st28gm06n052 is about two orders of magnitude, leading to much less efficient grain growth. Despite a higher seed-particle abundance in model st28gm05n033, the final degree of condensation is lower by about a factor of three, compared to the cooler, less massive model st28gm06n052, and the mass-loss rate is lower by about two orders of magnitude, corresponding to the difference in atmospheric gas densities. 

While the lower mass-loss rate of model st28gm05n033 can be explained by the lower atmospheric densities, it should also be mentioned here that the outflow of this hotter, more massive model is much more variable, due to the low efficiency of grain growth. As is shown in Sect.\,\ref{s:results_timevar}, most of the mass loss in model st28gm05n033 occurs during a few strong events ("gusts" of wind). Therefore, the mean structures in Fig.\,\ref{f:st28gm06n045_TimeAvgx} are less well defined than in the case of model st28gm06n052. 

This becomes apparent when taking a closer look at the mean radial-velocity curves. Model st28gm06n052 (red curve) shows the familiar pattern of a steady dust-driven wind: a steep initial increase in radial velocity associated with the dust formation region, followed by a levelling off further out, where both the radiative pressure and gravitational attraction weaken with increasing distance from the star. In contrast, the blue curve representing model st28gm05n033 shows a non-monotonic pattern, following a steep initial increase and a local maximum at about $1700\,R_\sun$. This point corresponds to the radius of the largest sphere fitting within the computational box (half of the edge length of the cubical box $x_\mathrm{outerbox}$, see Table\,\ref{t:ModelParam}). Beyond that point, the spatial averaging relies on the decreasing fractions of spherical surfaces contained in the box, until the largest distance, corresponding to $1.5\times x_\mathrm{outerbox}/2$, is reached. Computing representative averages for a gusty, inhomogeneous wind turns into a sampling problem, as the coverage of angles decreases. As a consequence, the average wind velocity of model st28gm05n033 is not well constrained. However, there is no doubt that material reaching the outer edge of the computational box will escape from the star, even if the average flow velocity does not exceed the local escape velocity (Fig.\,\ref{f:st28gm06n045_TimeAvgx}, dotted blue line). We note that the dynamics is not ballistic, and radiative acceleration continues to exceed gravitational attraction, while both forces decrease with distance. 

Figure~\ref{f:st28gm05n033_0375208_QuSeq1} shows evolving structures in model st28gm05n033 at three instants, for slices through the center of the star. The less efficient dust condensation is illustrated by the grain radius plots (bottom row), with only one pronounced dust formation event, followed by outward acceleration, taking place in the bottom right quadrant. In contrast to the corresponding Fig.~\ref{f:st28gm06n052_0233896_QuSeq1} for model st28gm06n052, the outer parts of model st28gm05n033 mostly show a red color in the radial-velocity plots during the chosen time interval, indicating gas falling back toward the star, after the passing of shock waves. Since the computational box of model st28gm05n033 is smaller by about a factor of two, this region should be compared to the corresponding parts of model st28gm06n052, which also show large regions of infalling gas. The somewhat clumpy but continuous outflow of model st28gm06n052 (blue colors in the radial-velocity plots) is more clearly visible at larger distances, beyond about 2000 solar radii.

\begin{figure*}[hbtp]
\begin{center}
\includegraphics[width=16.0cm]{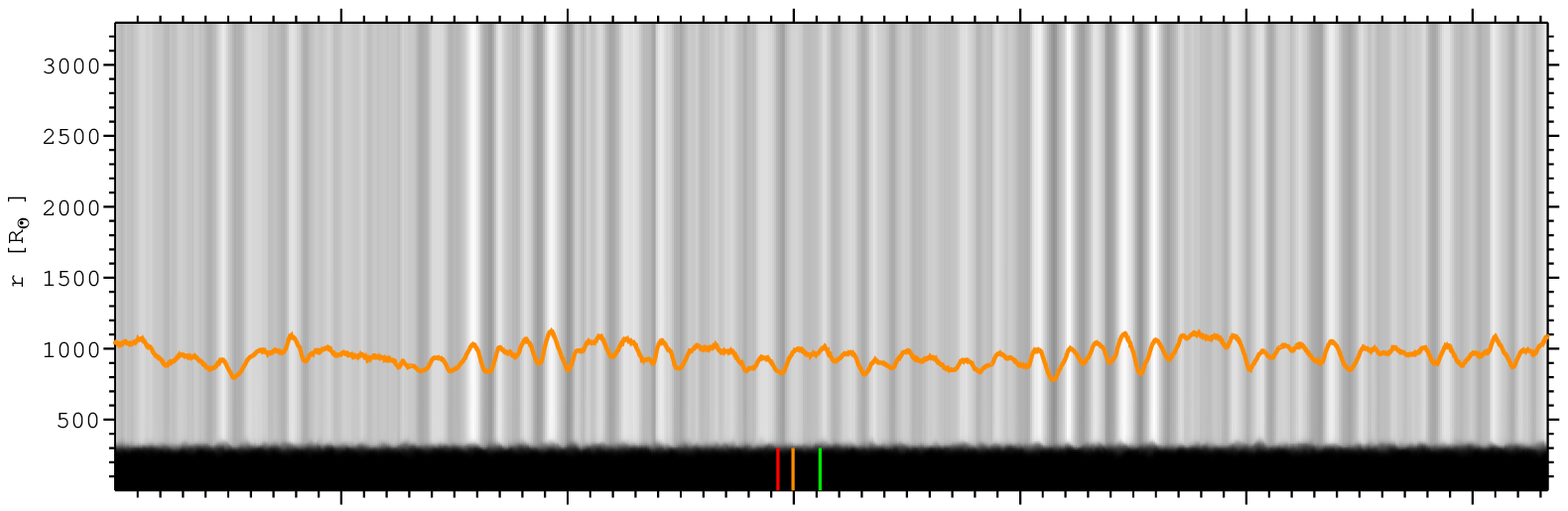}\includegraphics[width=2.0cm]{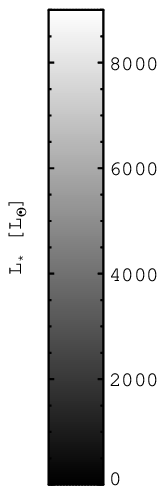}
\includegraphics[width=16.0cm]{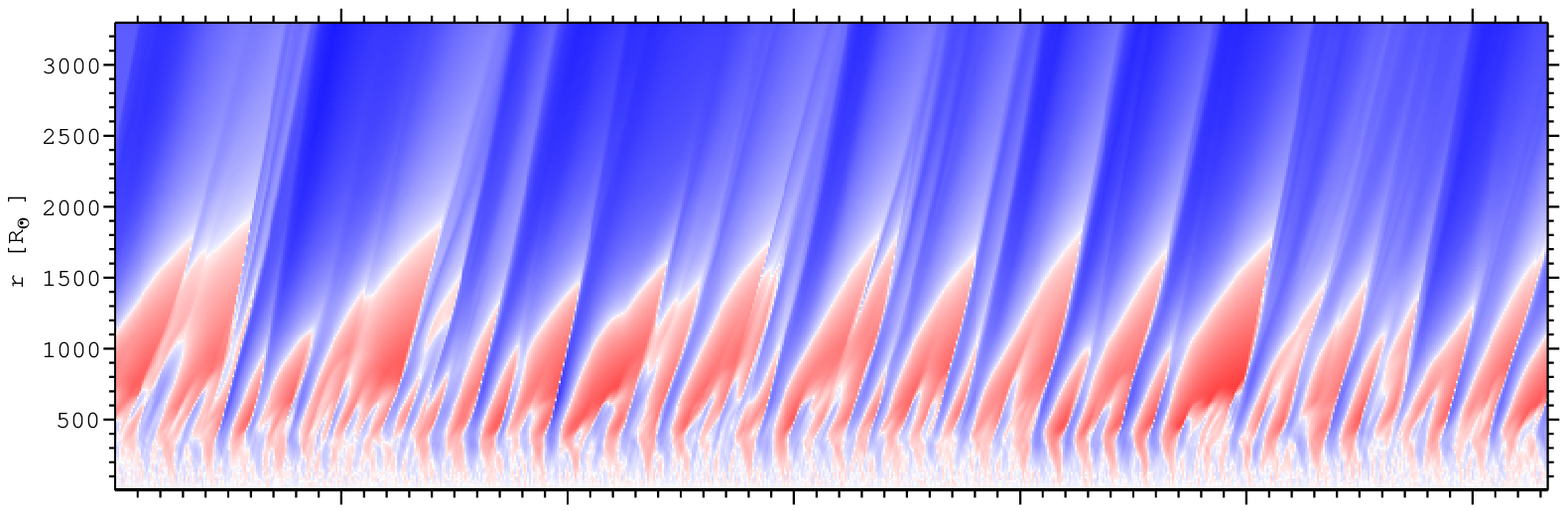}\includegraphics[width=2.0cm]{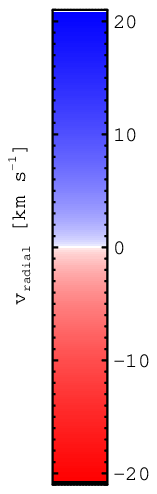}
\includegraphics[width=16.0cm]{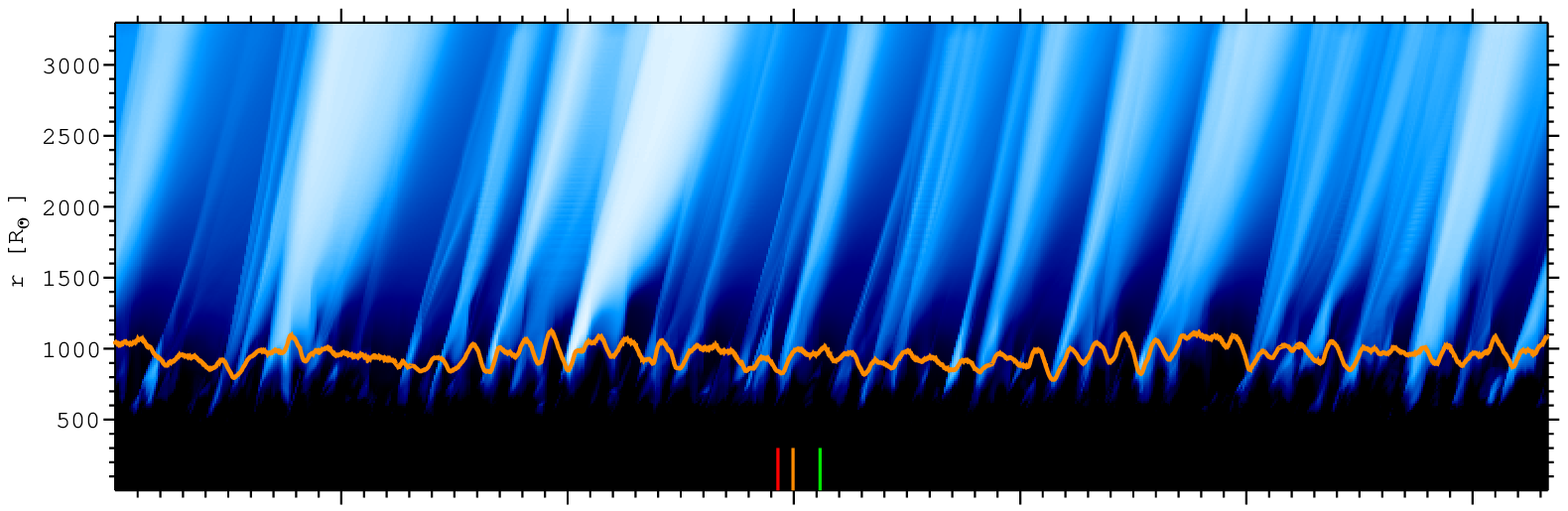}\includegraphics[width=2.0cm]{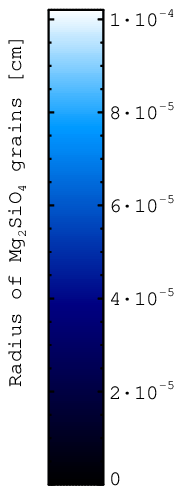}\vspace{0.4cm}
\end{center}
\caption{Spherical averages of luminosity (integrated radiative flux), radial velocity, and grain radius 
in the large 1\,$M_\sun$~model st28gm06n052, as a function of radial distance and time.
In the top and bottom panels, an isotherm with a temperature representative of silicate condensation is shown (1150\,K) to demonstrate the interplay between variable radiative heating and the inner edges of dust layers.
The red, orange, and green vertical bars indicate the instants selected
for Figs.\,\ref{f:st28gm06n052_0233896_QuSeq1} and \ref{f:st28gm06n052_0233896_IntSeq}.
The dark area in the lower part of the top panel represents the deep stellar interior, where most of the energy flux is transported by convection.
\label{f:st28gm06n052_QuOvertimeAndx}}
\end{figure*}

\begin{figure*}[hbtp]
\begin{center}
\includegraphics[width=16.0cm]{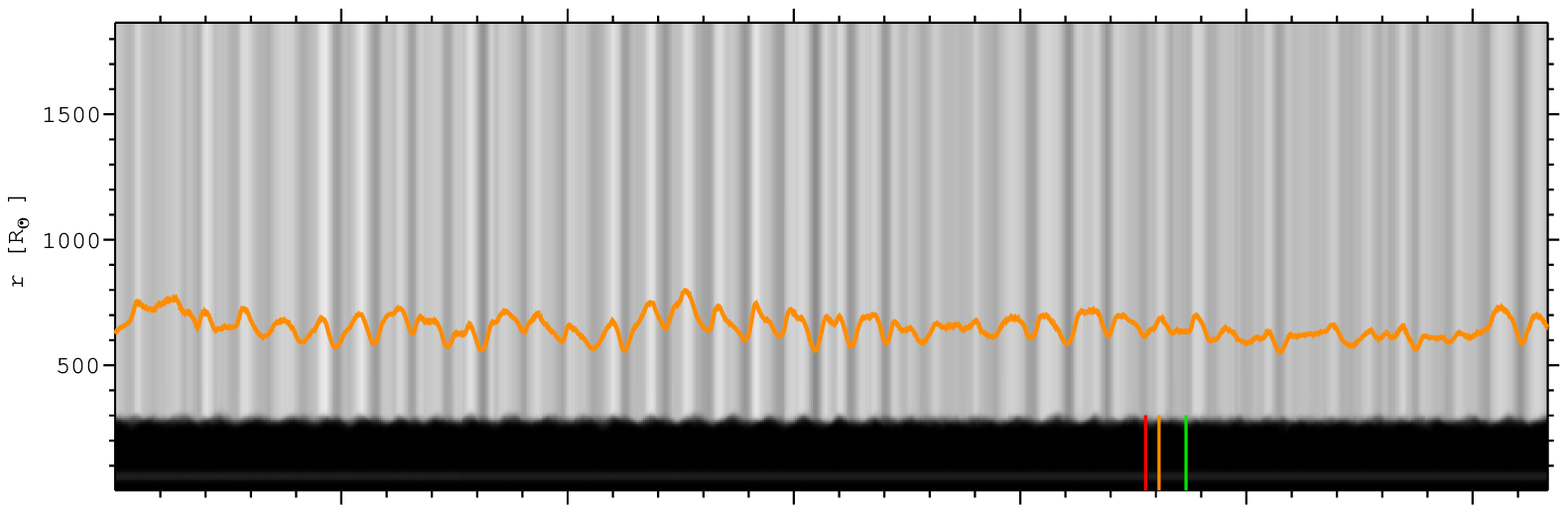}\includegraphics[width=2.0cm]{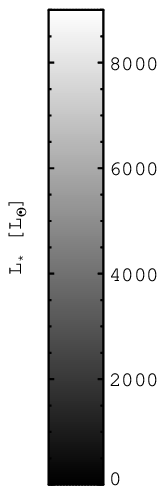}
\includegraphics[width=16.0cm]{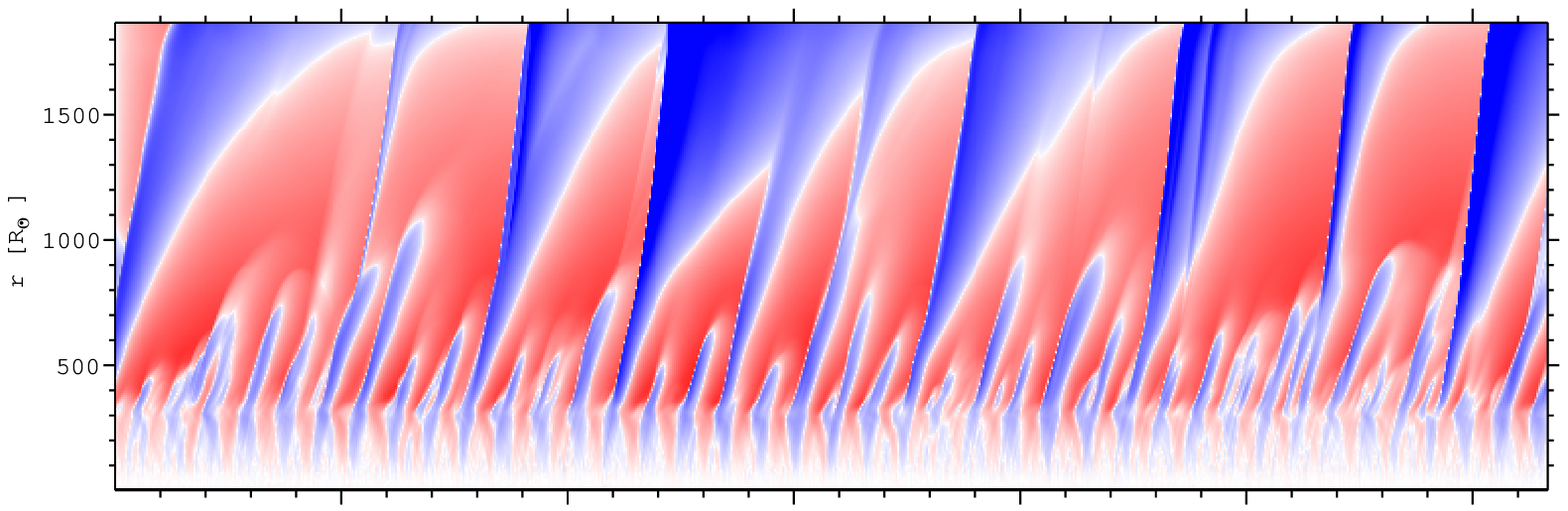}\includegraphics[width=2.0cm]{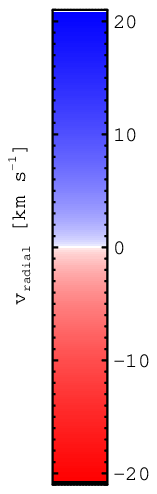}
\includegraphics[width=16.0cm]{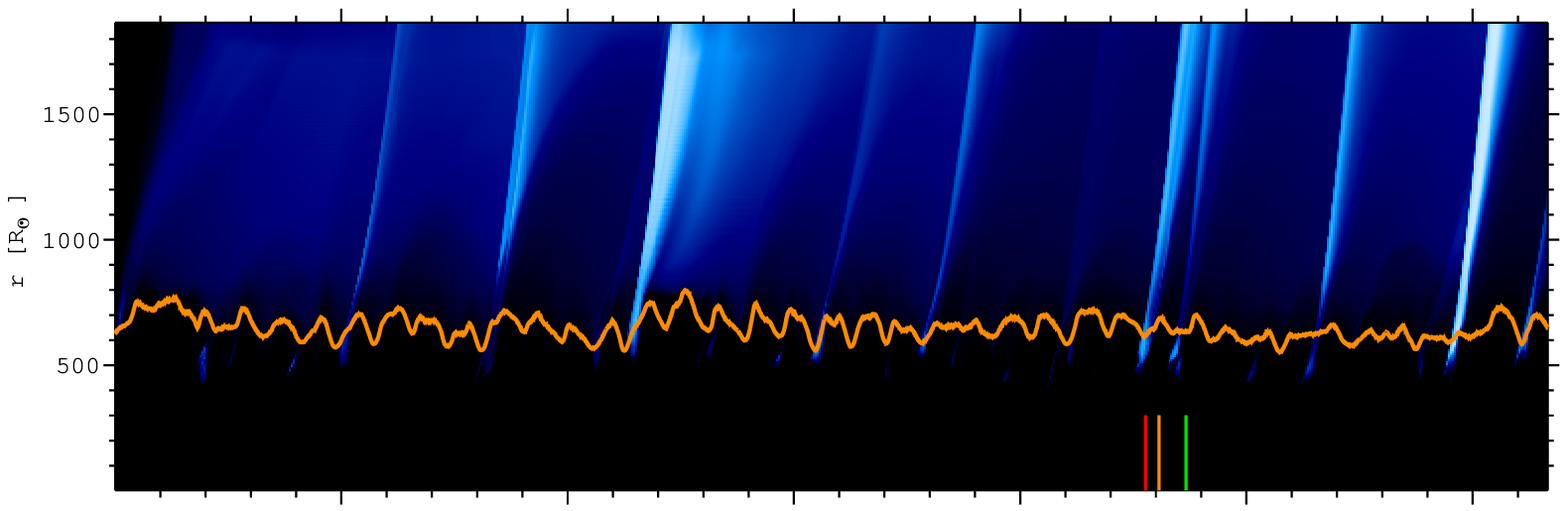}\includegraphics[width=2.0cm]{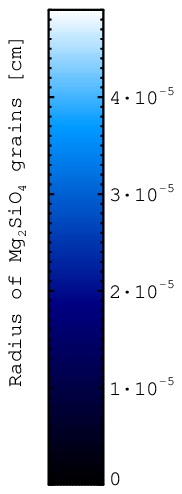}\vspace{0.4cm}
\end{center}
\caption{Spherical averages of luminosity (integrated radiative flux), radial velocity, and grain radius in the 1.5\,$M_\sun$~model st28gm05n033, as a function of radial distance and time.
In the top and bottom panels, an isotherm with a temperature representative of silicate condensation is shown (1150\,K) to demonstrate the interplay between variable radiative heating and the inner edges of dust layers.
The red, orange, and green vertical bars indicate the instants selected
for Figs.\,\ref{f:st28gm05n033_0375208_QuSeq1} and \ref{f:st28gm06n052_0233896_IntSeq}.
The dark area in the lower part of the top panel represents the deep stellar interior, where most of the energy flux is transported by convection.
The plots are similar to the corresponding ones in Fig\,\ref{f:st28gm06n052_QuOvertimeAndx}. However, it is important to note the differing spatial and temporal axes, and the different range in grain radii.
\label{f:st28gm05n033_QuOvertimeAndx}}
\end{figure*}

\subsection{Time variations of global properties}\label{s:results_timevar}

Figures showing the evolution of structures, as discussed above, are essential for understanding the interplay of physical processes at work. However, to study the connection between radial pulsation, dust formation, and wind acceleration, and to see variations in global properties as derived from spatially unresolved observations of AGB stars, we need a different approach. 

Figure~\ref{f:st28gm06n052_QuOvertimeAndx} shows spherical averages of luminosity, radial velocity, and grain radius as a function of radial depth and time for the cooler, less massive model st28gm06n052. 
In the panel showing luminosity, we can distinguish two different regions: the dark area (indicating low radiative flux) below about $300\,R_{\odot}$ represents the deep stellar interior, where essentially all of the energy flux is carried by convection. Further out, in the atmosphere and circumstellar envelope, on the other hand, energy is mostly transported by the radiative flux, which shows more or less regular variations related to radial pulsations of the star. 

In the velocity panel, the radial pulsations are apparent as alternating blue and red vertical stripes in the stellar interior. In the inner atmosphere, the dynamics is dominated by strong pulsation-induced shocks that are propagating outward (dark blue stripes, pointing upward and to the right), separated by phases where gas falls back toward the stellar surface (red). The outermost part of the model corresponds to the dust-driven stellar wind, with an outward-directed flow velocity (blue). We note that the highest wind velocities (darkest blue colors) occur in the wake of strong shocks, which, in turn, are directly linked to stellar expansion phases due to the pulsations. 

The temperatures in the atmosphere and wind-acceleration region are mostly set by radiative processes, dominated by photons emitted from the stellar surface. As the total radiative flux varies due to pulsations, so does the temperature in the dust-formation region. This effect is illustrated by an over-plotted isotherm (1150 K, typical silicate condensation temperature). During the luminosity minima, the atmosphere is coolest and the isotherm is closer to the center, meaning that the average distance from the star at which dust condensation is possible, is smallest. Since the density decreases steeply with the radial distance, the phases around luminosity minima are favorable for efficient dust condensation. This is apparent in the panel showing the grain radii, where pronounced dust formation events (bright areas) tend to coincide with deep minima in luminosity. 

A closer inspection shows that the regions of efficient grain growth (resulting in large grain radii) 
are typically closer to the star than the isothermal line, which roughly marks the condensation temperature of silicates. It should be pointed out that the radial distance of this line corresponds to a spherical mean, while the actual 2D isotherms in the 3D models are far from spherical. Figures~\ref{f:st28gm06n052_0233896_QuSeq1} and \ref{f:st28gm05n033_0375208_QuSeq1} illustrate that efficient grain growth events tend to occur in high-density regions where lower temperatures prevail close to the star. 

Figure~\ref{f:st28gm05n033_QuOvertimeAndx} shows spherical averages of luminosity, radial velocity, and grain radius as a function of radial depth and time for the warmer, more massive model st28gm05n033. In the radial-velocity plot, the alternating blue and red vertical stripes below about $300\,R_{\odot}$ in the stellar interior indicate radial pulsation, as discussed above for the other model. The velocity pattern both inside the star and in the inner atmosphere (propagating shocks, indicated by blue stripes pointing upward and to the right) looks somewhat more regular for this star with higher surface gravity. However, fewer of the global-scale atmospheric shock waves lead to efficient dust formation events (see the grain-radius panel), followed by outward acceleration of the gas-dust mixture. A majority of the pulsation-induced shocks are stopped by infalling material (red). We note, again, that the computational box is smaller than for model st28gm06n052, and the global flow direction in the outer parts of the model is not always directed outward (both blue and red colors can be found at the upper edge). This is consistent with the less efficient dust formation and less dense outflow, as indicated by the spherically and temporally averaged structures shown in Fig.~\ref{f:st28gm06n045_TimeAvgx}.

\begin{figure*}[hbtp]
\begin{center}
\hspace*{0.9cm}\includegraphics[width=15.3cm]{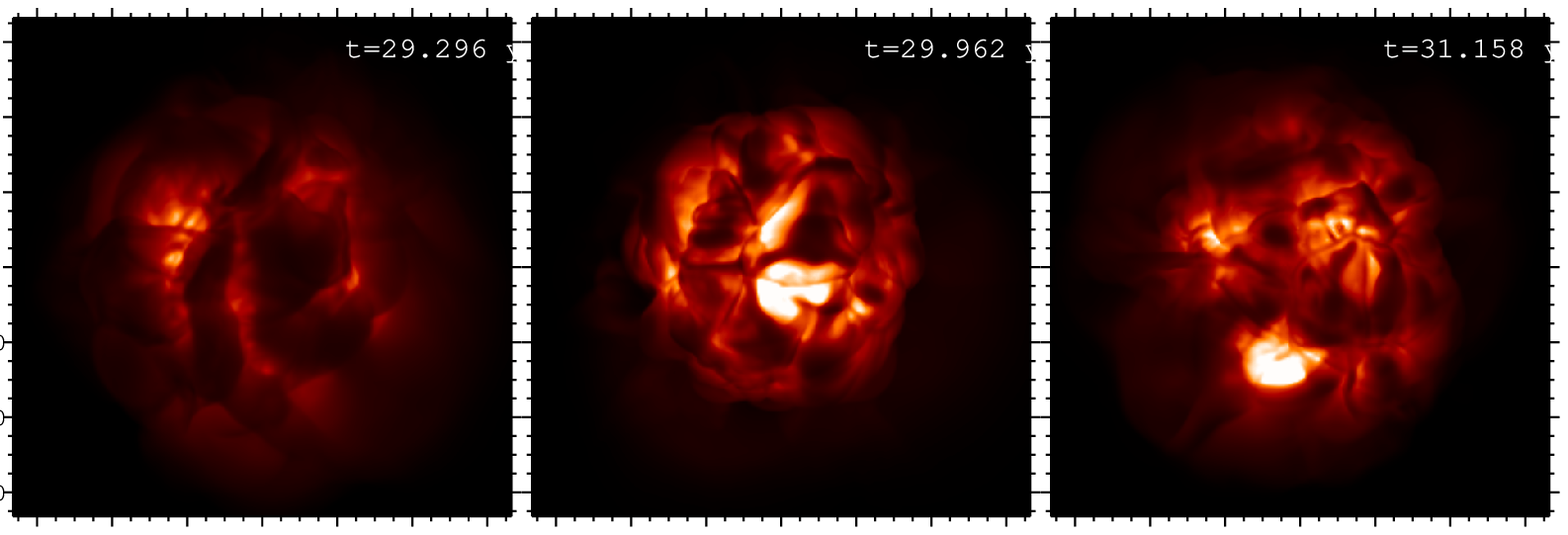}\includegraphics[width=1.9125cm]{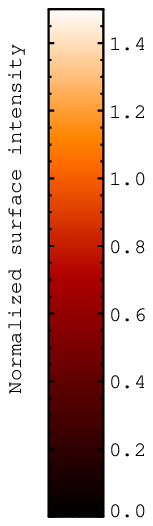}\vspace*{0.7cm}
\hspace*{0.9cm}\includegraphics[width=15.3cm]{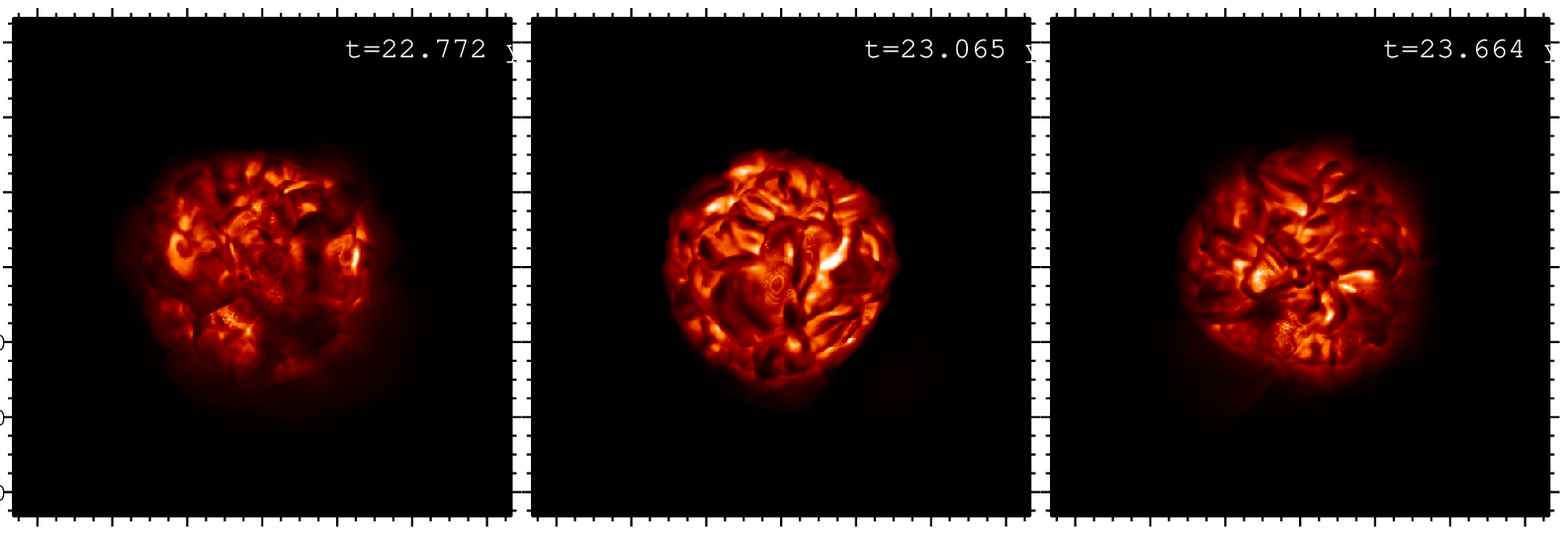}\includegraphics[width=1.9125cm]{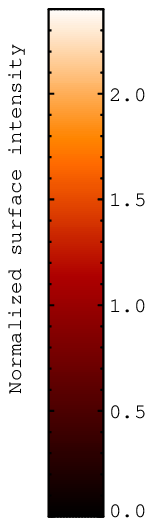}\vspace{0.3cm}
\end{center}
\caption{Time sequences of bolometric surface intensity
  for the large 1\,$M_\sun$~model st28gm06n052
    (snapshots about 8 and 14 months apart, see the counter at the top of the panels)
  and the 1.5\,$M_\sun$~model st28gm05n033
    (snapshots about 3.5 and 7 months apart).
  Only the inner part of computational box that contains the star is plotted,
  on the same scale for both models.
\label{f:st28gm06n052_0233896_IntSeq}}
\end{figure*}

\section{Discussion}\label{s:discussion}

\subsection{Comparison with earlier 3D models and with 1D models}

Dynamical models and spatially resolved observations indicate that the extended 
atmospheres of AGB stars are characterized by a network of shock waves, triggered by large-scale convective flows and stellar pulsations. Our earlier 3D AGB-star models suggested that the resulting structures in atmospheric densities should lead to a patchy dust distribution in the circumstellar material close to the star \citep{Hoefner2019A&A...623A.158H}.\footnote{For clarity, we point out that this 
effect is different from the intrinsic instabilities of a circumstellar dust shell, as discussed by \cite{Woitke2006A&A...452..537W}, occurring in 2D models that did not take processes in the star and their 
influences on the circumstellar environment into account.} 
While providing an explanation for the origin of clumpy dust clouds observed around several nearby AGB stars, these earlier models did not include radiation pressure, and we could therefore not draw definite conclusions about atmospheric dynamics and the formation of a dust-driven stellar wind. 

In contrast, the new models presented in Sect.~\ref{s:results} account for the effects of radiation pressure on the silicate dust that forms in the stellar atmosphere, with grain growth rates depending on the densities of condensible material in the surrounding gas. Dust condensation is most efficient in the dense wakes of shock waves, and the basic cloud formation mechanism, which precedes wind acceleration, is similar to our earlier 3D models. However, we note that the models presented here show more pronounced 
spatial variations in grain size, and the final degree of condensation in the outflow, when averaging over time, is far from complete (Table\,\ref{t:mod}). 

Incomplete condensation is in line with results of 1D dust-driven wind models, where grain growth slows down drastically in the accelerating outflow due to rapidly falling densities.  This typically leaves a significant fraction of condensible material in the gas phase \citep[see, e.g.,][]{Hoefner2016A&A...594A.108H, Hoefner2022A&A...657A.109H}. Another feature of the new 3D models that is reminiscent of 1D dust-driven wind models is the time-dependent behavior of spherically averaged quantities, as seen in Figs.\,\ref{f:st28gm06n052_QuOvertimeAndx} and \ref{f:st28gm05n033_QuOvertimeAndx}. The spherical averaging highlights the role of radial stellar pulsations, related luminosity variations, and resulting large-scale atmospheric shock waves, which trigger dust formation events. 

As temperatures and gas densities vary along a shock front,
in contrast to the situation in 1D models,
the dust forms with different rates at different locations, leading to regions with very high, and others with low or zero, dust densities
behind a shock.
Depending on the statistics of the thermodynamic quantities, the net effect on
the overall mass loss might be an increase or a decrease,
compared to 1D models.
The stellar parameter sets of the 3D models presented here were chosen to fall into two different regimes: according to results from 1D DARWIN simulations, model st28gm06n052 is expected to develop a pronounced dust-driven wind, while the 1D counterpart of model st28gm05n033 fails to produce an outflow \citep[see][]{Bladh2019A&A...626A.100B}. In this context it is remarkable that the 1.5\,$M_\sun$~CO5BOLD model does develop a gusty, inhomogeneous wind with a low mass-loss rate, in contrast to its wind-less DARWIN equivalent. This could be due to an intrinsically more efficient wind mechanism in the 3D models, that is to say, more favorable conditions for dust formation in the inhomogeneous atmospheres, with cool dense pockets of gas close to the star. However, there are differences in the treatment of radiative transfer compared to the DARWIN models that may also affect the results and need to be investigated further, before more quantitative conclusions can be drawn (see Sect.\,\ref{s:approx}).

\begin{figure*}[hbtp]
\begin{center}
\hspace*{0.9cm}\includegraphics[width=15.3cm]{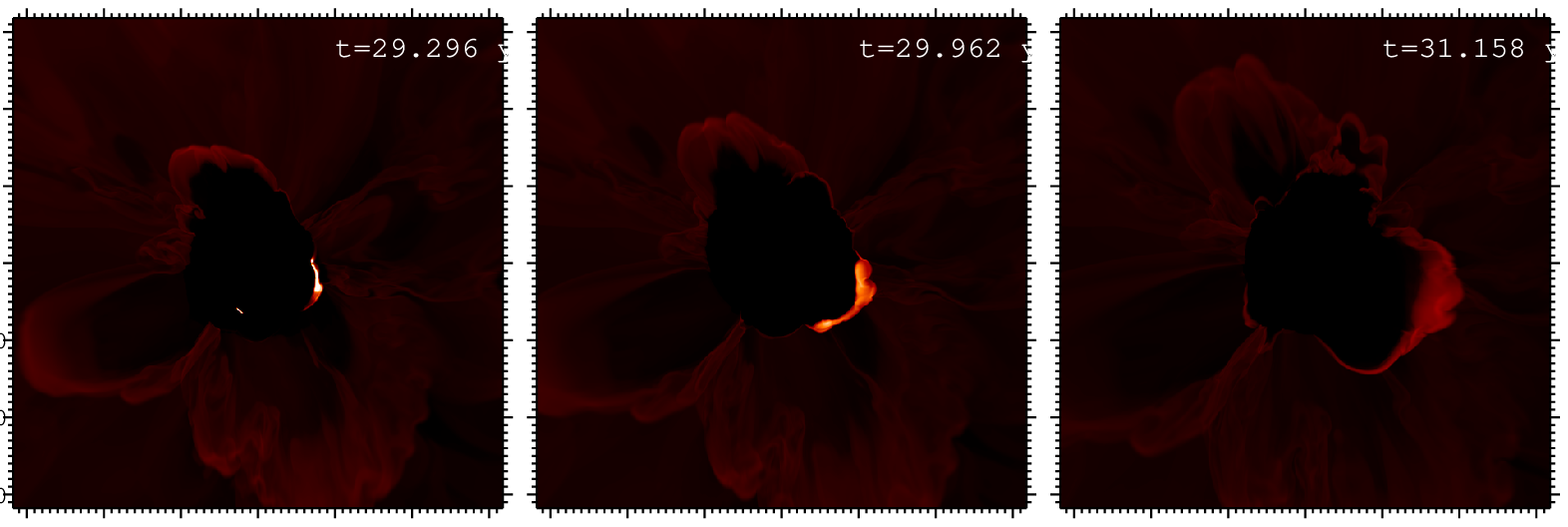}\includegraphics[width=1.9cm]{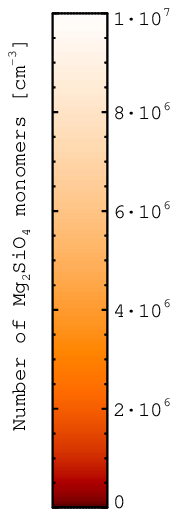}\vspace{0.3cm}
\end{center}
\caption{Time sequence of silicate density,
  through the center of the large 1\,$M_\sun$~model st28gm06n052.
  The snapshots are about 8 and 14 months apart, respectively
 (see the counter at the top of the panels).
\label{f:st28gm06n052_0233896_QuSeq2}}
\end{figure*}

\subsection{Observable properties}

Using averages of radial velocities and other quantities, taken over spherical shells, \cite{Freytag2017A&A...600A.137F} identified radial pulsation in a sample of 3D star-in-a-box models with different stellar parameters. The simulations show periods in good agreement with the observed $P$-$L$ relation for Mira variables by \cite{Whitelock2009MNRAS.394..795W}. Recently, \citet{Ahmad2023A&A...669A..49A}
analyzed the pulsation properties of a much larger set of global 3D models produced with the CO5BOLD code, including those presented here. The periods determined for models st28gm06n052 and st28gm05n033 are 545 and 297 days, respectively, bracketing the mean $P$-$L$ relation derived by \cite{Whitelock2009MNRAS.394..795W}, with both models falling inside the broad range of observed values. 

While the spherically averaged radial velocities show clear signs of radial stellar pulsations and  dominantly outward-directed flow velocities in the outer parts of the models (see Figs.\,\ref{f:st28gm06n052_QuOvertimeAndx} and \ref{f:st28gm05n033_QuOvertimeAndx}), the 2D slices demonstrate the complex nature of dynamics in the convective stellar interior and in the stellar atmosphere (Figs.\,\ref{f:st28gm06n052_0233896_QuSeq1} and \ref{f:st28gm05n033_0375208_QuSeq1}). Convective motions, and possibly non-radial pulsations, cause shock waves on a wide range of angular scales. The shocks may merge or collide, leading to intricate density structures. Some shock waves trigger sufficient dust formation and radiative acceleration to overcome gravity, leading to an outflow. At the same time, other parts of the atmosphere show ballistic motions and material falling back toward the stellar surface. Both the density and the temperature in the atmosphere are strongly variable, and differ from the spherical structures expected in case of a purely radial pulsation. Recent submillimeter observations have given evidence of simultaneous inward and outward directed motions of atmospheric gas, and a complex morphology of the extended dynamical atmospheres of AGB stars, with coexisting warm and cool gas components \citep[e.g.,][]{Khouri2016MNRAS.463L..74K, Vlemmings2017NatAs...1..848V}. 

\citet[][]{Paladini2018Natur.553..310P} presented H-band images of {$\pi$}$^{1}$ Gruis (based on VLTI/PIONIER data), indicating large granulation cells on the stellar surface. The sizes of the observed surface structures are in good agreement with extrapolations of local 3D models for less evolved stars, and with qualitative predictions of global 3D AGB-star models. Figure\,\ref{f:st28gm06n052_0233896_IntSeq} shows time sequences of the bolometric surface intensity for 
the large 1\,$M_\sun$~model 
and the 1.5\,$M_\sun$~model 
plotted on the same spatial scale (zoomed in on the star, which is small compared to the computational box). Both models have a similar mean luminosity, but the higher mass in model st28gm05n033, together with a slightly smaller radius, leads to a more well-defined stellar surface with much smaller structures. The lower surface gravity in model st28gm06n052, on the other hand, results in a much more extended atmosphere, favoring efficient dust formation. It should be noted here that the images show the bolometric intensity. For a detailed comparison with observations, images in appropriate filter bands have to be computed \cite[see][and references therein]{Chiavassa2018A&A...617L...1C}, which is an effort well beyond the scope of the present paper. However, the qualitative trends with stellar parameters can be expected to be similar to the bolometric images. 

Dust condensation is favored by high density, and therefore grain growth is most effective in the dense wakes of outward-propagating shocks, as illustrated by the sequence of grain-size plots in Fig.\,\ref{f:st28gm06n052_0233896_QuSeq1} for model st28gm06n052. Figure\,\ref{f:st28gm06n052_0233896_QuSeq2} shows the corresponding plots of dust density (total number of monomers contained in dust grains per unit volume). This quantity can be regarded as a simple proxy for dust emission in the mid-IR, where dust grains are small compared to the wavelengths, and the dust opacity (emission coefficient) is roughly proportional to the volume occupied by the dust.
The formation of a new dust cloud in the wake of a shock, followed by radiative acceleration away from the star, is clearly visible in the lower right quadrant. In contrast to the grain size, however, the density decreases as the cloud moves outward, as indicated by the darkening of the colors.

\subsection{Approximations and future improvements}\label{s:approx}

In current 1D DARWIN models, frequency-dependent radiative transfer is routinely computed at several hundred wavelength points,  thereby achieving a good coverage of the spectral energy distribution, while simultaneously solving the equations of hydrodynamics and dust formation \citep[see][]{Hoefner2016A&A...594A.108H,Hoefner2022A&A...657A.109H}. The much more computationally intensive 3D AGB-star models, however, are based on an opacity binning scheme, to account for non-gray effects (see Sect.\,\ref{s:box_radtra}). In this context, it is a nontrivial task to introduce dust opacities, which are not simply a function of current temperature and density as the pre-tabulated gas opacities are, but depend on grain sizes that result from nonequilibrium condensation and evaporation processes, and are not known a priori. 

In the exploratory 3D "star-and-wind-in-a-box" models presented here, we chose to represent the dust opacities relevant for radiation pressure by monochromatic values at a wavelength near the stellar flux maximum. Furthermore, considering the level of other approximations in the radiative transfer, we describe the size-dependent efficiency factor for radiation pressure on dust grains by a simple analytical formula (see Sect.\,\ref{s:dust_rp}). The formula reproduces key features of Fe-free silicate dust, such as low absorption in the visual and near-IR, and a strong dependence of scattering on grain size for particles with radii smaller than the relevant wavelengths. It should give a qualitatively correct picture of the dynamics, and the resulting wind velocities are, indeed, in the expected range. Nevertheless, we plan to replace the simple formula with optical properties computed with Mie theory in future 3D models, similar to what is used in the 1D DARWIN models. 

Another obvious difference compared to 1D wind models is the smaller overall size of the computational domain, adjusted for an efficient use of computational resources. For model st28gm06n052, the box extends to distances where the stellar wind is well established and close to reaching its terminal velocity. The  more massive, somewhat hotter model st28gm05n033, with its smaller box, on the other hand, illustrates potential problems with determining reliable average wind properties for stars with higher surface gravity and less favorable conditions for dust formation, leading to strongly inhomogeneous outflows with low mass-loss rates. In principle, the box of this particular model could have been extended to larger distances. In practice, however, the model in its current form is sufficient to demonstrate the influence of key stellar parameters on the formation of a dust-driven wind.

\section{Conclusions}\label{s:conclusions}

In this paper we have presented the first global 3D RHD models of AGB stars and their dust-driven winds, computed with the CO5BOLD code. The models allow us to follow the flow of matter from the stellar interior, through the dynamical atmosphere, and into the wind-acceleration region, while taking nonequilibrium dust formation and the interaction of matter with radiation into account. Convection and pulsations emerge self-consistently in the 3D models. Wind properties (e.g. mass-loss rates and outflow velocities) can be derived without relying on parameterized descriptions of these interior dynamical processes, in contrast to current 1D wind models. Atmospheric shock waves, triggered by convective motions and pulsation, play a critical role in the condensation of the dust grains, which drive the winds. A global 3D approach is essential to make progress in understanding dynamical processes in AGB stars, and to solve long-standing problems regarding mass loss. 

The giant convection cells, which are a characteristic feature of AGB stars, cause large-scale intensity patterns in the stellar photosphere, and leave their imprints on atmospheric temperatures and densities. 
In contrast to 1D models with purely radial pulsations and global spherical shocks fronts propagating outward through the atmosphere, the shocks induced by convection and pulsation in the 3D models cover large but finite regions. This leads to a patchy, nonspherical structure of the atmosphere. Since the efficiency of dust condensation depends critically on gas density, new dust clouds mostly emerge in the dense wakes of atmospheric shocks. The resulting clumpy distribution of newly formed dust, together with radiative pressure being highly sensitive to grain sizes, leads to a complex 3D morphology of the extended atmosphere and wind-acceleration zone, with simultaneous infall and outflow regions close to the star. 

Genuine 3D effects in the models are strong deviations of isotherms from spherical symmetry and the intermittent formation of cool pockets of gas in the stellar vicinity.
Efficient dust formation tends to occur closer to the star than spherical averages of the temperature would indicate, in dense regions where grain growth rates are higher than average. 
This can lead to dust-driven outflows with low mass-loss rates in situations where 1D models with the same stellar parameters do not produce winds. 
In contrast, for stars where the overall conditions for dust formation and wind acceleration are favorable, it is not obvious whether the resulting mass-loss rates will be higher or lower than in the 1D case. 
The increased efficiency of dust formation in high-density clumps may be set off by a low filling factor of these regions.

The key features of the models (high-contrast convective surface patterns, complex velocity fields, clumpy dust clouds) are seen in observations of nearby AGB stars. Synthetic observables need to be computed from the dynamical 3D structures in order to compare the models to observations in more detail. The production of spectra and images in various wavelength regimes is under way, but this is a considerable, time-consuming effort, and results will be presented in future papers. 

Based on the first exploratory models discussed in this paper, we can mainly draw qualitative conclusions about the physics and 3D morphology of AGB-star atmospheres and winds, since the simulations include a number of approximations. Improvements regarding dust microphysics and the treatment of radiation pressure are planned for future models, to get a better quantitative description of these aspects, and of the resulting wind properties.  

Eventually, it will be necessary to produce additional models, to explore how mass-loss rates depend on fundamental stellar parameters. However, 3D models are time-consuming to compute and analyze. Therefore, 1D models of dust-driven winds will continue to play an important role for the foreseeable future, as they can provide the wide coverage of stellar parameter combinations required in stellar evolution models. Hopefully, knowledge gained from full 3D modeling can be used to improve 1D models, in particular regarding a better representation of pulsation and convection effects on the stellar atmosphere and wind.

\begin{acknowledgements}

This work is part of a project that has received funding from 
the European Research Council (ERC) 
under the European Union’s Horizon 2020 research and innovation programme 
(Grant agreement No.~883867, project EXWINGS) 
and the Swedish Research Council ({\it Vetenskapsrådet}, grant number 2019-04059). 
The computations were enabled by resources provided by the 
Swedish National Infrastructure for Computing (SNIC)
partially funded by the Swedish Research Council through grant agreement no.~2018-05973.
We thank the referee Jan Martin Winters for his insightful comments.

\end{acknowledgements}

\bibliographystyle{aa}    
\bibliography{aa_redsg}

\end{document}